\documentclass[manuscript]{aastex61}
\usepackage[hyphenbreaks]{breakurl}
\usepackage[latin1]{inputenc}
\submitjournal{ApJ}

\shorttitle{Forbush decreases and $<$ 2-day GCR flux variations}
\shortauthors{Armano et al.}

\begin{document}

\title {Forbush decreases and $<$ 2-day GCR flux non-recurrent variations studied with LISA Pathfinder}

\correspondingauthor{Catia Grimani}
\email{catia.grimani@uniurb.it}

\author{M. Armano}
\affiliation{European Space Astronomy Centre, European Space Agency \\
Villanueva de la Ca\~{n}ada\\
28692 Madrid, Spain}

\author{H. Audley}
\affiliation{Albert-Einstein-Institut, Max-Planck-Institut f\"ur Gravitationsphysik und Leibniz Universit\"at Hannover \\
Callinstra{\ss}e 38 \\
30167 Hannover, Germany}

\author{J. Baird}
\affiliation{High Energy Physics Group, Physics Department, Imperial College London \\
Blackett Laboratory, Prince Consort Road \\
London, SW7 2BW, UK}


\author{S. Benella}
\affil{DISPEA, Universit\`a di Urbino ``Carlo Bo'' \\Via S. Chiara, 27 \\61029 Urbino, Italy}
\affil{INFN - Sezione di Firenze\\ via G. Sansone 1\\50019, Sesto Fiorentino, Firenze, Italy}

\author{P. Binetruy}
\altaffiliation{Deceased 1 April 2017}
\affiliation{APC, Univ Paris Diderot, CNRS/IN2P3, CEA/lrfu, Obs de Paris, Sorbonne Paris Cit\'e, France}

\author{M. Born}
\affiliation{Albert-Einstein-Institut, Max-Planck-Institut f\"ur Gravitationsphysik und Leibniz Universit\"at Hannover \\
Callinstra{\ss}e 38 \\
30167 Hannover, Germany}

\author{D. Bortoluzzi}
\affiliation{Department of Industrial Engineering, University of Trento, via Sommarive 9, 38123 Trento, and Trento Institute for Fundamental Physics and Application / INFN}

\author{E. Castelli}
\affiliation{Dipartimento di Fisica, Universit\`a di Trento and Trento Institute for Fundamental Physics and Application  
INFN\\
38123 Povo, Trento, Italy}

\author{A. Cavalleri}
\affiliation{Istituto di Fotonica e Nanotecnologie, CNR-Fondazione Bruno Kessler, I-38123 Povo, Trento, Italy}

\author{A. Cesarini}
\affil{DISPEA, Universit\`a di Urbino ``Carlo Bo'' \\Via S. Chiara, 27 \\61029 Urbino, Italy}

\author{A. M. Cruise}
\affiliation{The School of Physics and Astronomy, University of Birmingham\\ Birmingham, UK}

\author{K. Danzmann}
\affiliation{Albert-Einstein-Institut, Max-Planck-Institut f\"ur Gravitationsphysik und Leibniz Universit\"at Hannover \\
Callinstra{\ss}e 38 \\
30167 Hannover, Germany}

\author{M. de Deus Silva}
\affiliation{European Space Astronomy Centre, European Space Agency \\
Villanueva de la Ca\~{n}ada\\
28692 Madrid, Spain}

\author{I. Diepholz}
\affiliation{Albert-Einstein-Institut, Max-Planck-Institut f\"ur Gravitationsphysik und Leibniz Universit\"at Hannover \\
Callinstra{\ss}e 38 \\
30167 Hannover, Germany}

\author{G. Dixon}
\affiliation{The School of Physics and Astronomy, University of Birmingham\\ Birmingham, UK}

\author{R. Dolesi}
\affiliation{Dipartimento di Fisica, Universit\`a di Trento and Trento Institute for Fundamental Physics and Application / INFN\\ 38123 Povo, Trento, Italy}

\author{M. Fabi}
\affil{DISPEA, Universit\`a di Urbino ``Carlo Bo'' \\Via S. Chiara, 27 \\61029 Urbino, Italy}

\author{L. Ferraioli}
\affiliation{Institut f\"ur Geophysik, ETH Z\"urich\\ Sonneggstrasse 5\\ CH-8092, Z\"urich, Switzerland}

\author{V. Ferroni}
\affiliation{Dipartimento di Fisica, Universit\`a di Trento and Trento Institute for Fundamental Physics and Application / INFN\\ 38123 Povo, Trento, Italy}

\author{N. Finetti}
\affiliation{INFN - Sezione di Firenze\\ via G. Sansone 1\\50019, Sesto Fiorentino, Firenze, Italy}
\affiliation{Dipartimento di Scienze Fisiche e Chimiche, Universit\`a degli Studi dell'Aquila\\ Via Vetoio, Coppito, \\67100 L'Aquila, Italy}

\author{E. D. Fitzsimons}
\affiliation{The UK Astronomy Technology Centre, Royal Observatory Edinburgh\\ Blackford Hill, Edinburgh\\ EH9 3HJ, UK}

\author{M. Freschi}
\affiliation{European Space Astronomy Centre, European Space Agency \\
Villanueva de la Ca\~{n}ada\\
28692 Madrid, Spain}

\author{L. Gesa}
\affiliation{
Institut de Ci\`encies de l'Espai (ICE, CSIC) Campus UAB, Carrer de Can Magrans s/n E-08193 Cerdanyola del Vall\`es
Institut d'Estudis Espacial de Catalunya (IEEC) Edifici Nexus I, C/ Gran Capit\`a 2-4, despatx 201 E-08034 Barcelona, Spain
}

\author{F. Gibert}
\affiliation{Dipartimento di Fisica, Universit\`a di Trento and Trento Institute for Fundamental Physics and Application / INFN\\ 38123 Povo, Trento, Italy}

\author{D. Giardini}
\affiliation{Institut f\"ur Geophysik, ETH Z\"urich\\ Sonneggstrasse 5\\ CH-8092, Z\"urich, Switzerland}

\author{R. Giusteri}
\affiliation{Dipartimento di Fisica, Universit\`a di Trento and Trento Institute for Fundamental Physics and Application / INFN\\ 38123 Povo, Trento, Italy}

\author[0000-0002-5467-6386]{C. Grimani}
\affiliation{DISPEA, Universit\`a di Urbino ``Carlo Bo'' \\Via S. Chiara, 27 \\61029 Urbino, Italy}
\affiliation{INFN - Sezione di Firenze\\ via G. Sansone 1\\50019, Sesto Fiorentino, Firenze, Italy}

\author{J. Grzymisch}
\affiliation{European Space Technology Centre, European Space Agency\\ Keplerlaan 1\\ 2200 AG Noordwijk, The Netherlands} 
\author{I. Harrison}
\affiliation{European Space Operations Centre\\ European Space Agency\\ 64293 Darmstadt, Germany}

\author{G. Heinzel} 
\affiliation{Albert-Einstein-Institut, Max-Planck-Institut f\"ur Gravitationsphysik und Leibniz Universit\"at Hannover \\
Callinstra{\ss}e 38 \\
30167 Hannover, Germany}

\author{M. Hewitson}
\affiliation{Albert-Einstein-Institut, Max-Planck-Institut f\"ur Gravitationsphysik und Leibniz Universit\"at Hannover \\
Callinstra{\ss}e 38 \\
30167 Hannover, Germany}

\author{D. Hollington}
\affiliation{High Energy Physics Group, Physics Department, Imperial College London \\
Blackett Laboratory, Prince Consort Road \\
London, SW7 2BW, UK}

\author{D. Hoyland}
\affiliation{The School of Physics and Astronomy, University of Birmingham\\ Birmingham, UK}

\author{M. Hueller}
\affiliation{Dipartimento di Fisica, Universit\`a di Trento and Trento Institute for Fundamental Physics and Application / INFN\\ 38123 Povo, Trento, Italy}

\author{H. Inchausp\'e}
\affiliation{APC, Univ Paris Diderot, CNRS/IN2P3, CEA/lrfu, Obs de Paris, Sorbonne Paris Cit\'e, France}

\author{O. Jennrich}
\affiliation{European Space Technology Centre, European Space Agency\\ Keplerlaan 1\\ 2200 AG Noordwijk, The Netherlands}

\author{P. Jetzer}
\affiliation{Physik Institut, Universit\"at Z\"urich\\ Winterthurerstrasse 190\\ CH-8057 Z\"urich, Switzerland}

\author{N. Karnesis}
\affiliation{Albert-Einstein-Institut, Max-Planck-Institut f\"ur Gravitationsphysik und Leibniz Universit\"at Hannover \\
Callinstra{\ss}e 38 \\
30167 Hannover, Germany}

\author{B. Kaune}
\affiliation{Albert-Einstein-Institut, Max-Planck-Institut f\"ur Gravitationsphysik und Leibniz Universit\"at Hannover \\
Callinstra{\ss}e 38 \\
30167 Hannover, Germany}

\author{N. Korsakova}
\affiliation{SUPA, Institute for Gravitational Research\\ School of Physics and Astronomy, University of Glasgow\\ Glasgow, G12 8QQ, UK}

\author{C. J. Killow}
\affiliation{SUPA, Institute for Gravitational Research\\ School of Physics and Astronomy, University of Glasgow\\ Glasgow, G12 8QQ, UK}

\author{K. Kudela}
\altaffiliation{Deceased 20 January 2019}
\affiliation{Nuclear Physics Institute of the CAS, \v{R}e\v{z}, Czech Republic}

\author{M. Laurenza}
\affiliation{INFN - Sezione di Firenze\\ via G. Sansone 1\\50019, Sesto Fiorentino, Firenze, Italy}
\affiliation{Istituto di Astrofisica e Planetologia Spaziali\\ INAF, Roma, Italy}

\author{J. A. Lobo}
\altaffiliation{Deceased 30 September 2012}
\affiliation{
Institut de Ci\`encies de l'Espai (ICE, CSIC) Campus UAB, Carrer de Can Magrans s/n E-08193 Cerdanyola del Vall\`es
Institut d'Estudis Espacial de Catalunya (IEEC) Edifici Nexus I, C/ Gran Capit\`a 2-4, despatx 201 E-08034 Barcelona, Spain
}

\author{I. Lloro}
\affiliation{
Institut de Ci\`encies de l'Espai (ICE, CSIC) Campus UAB, Carrer de Can Magrans s/n E-08193 Cerdanyola del Vall\`es
Institut d'Estudis Espacial de Catalunya (IEEC) Edifici Nexus I, C/ Gran Capit\`a 2-4, despatx 201 E-08034 Barcelona, Spain
}

\author{L. Liu}
\affiliation{Dipartimento di Fisica, Universit\`a di Trento and Trento Institute for Fundamental Physics and Application / INFN\\ 38123 Povo, Trento, Italy}

\author{J. P. L\'opez-Zaragoza}
\affiliation{
Institut de Ci\`encies de l'Espai (ICE, CSIC) Campus UAB, Carrer de Can Magrans s/n E-08193 Cerdanyola del Vall\`es
Institut d'Estudis Espacial de Catalunya (IEEC) Edifici Nexus I, C/ Gran Capit\`a 2-4, despatx 201 E-08034 Barcelona, Spain
}

\author{R. Maarschalkerweerd}
\affiliation{European Space Operations Centre\\ European Space Agency\\ 64293 Darmstadt, Germany}

\author{D. Mance}
\affiliation{Institut f\"ur Geophysik, ETH Z\"urich\\ Sonneggstrasse 5\\ CH-8092, Z\"urich, Switzerland}

\author{N. Meshksar}
\affiliation{Institut f\"ur Geophysik, ETH Z\"urich\\ Sonneggstrasse 5\\ CH-8092, Z\"urich, Switzerland}

\author{V. Mart\'in}
\affiliation{
Institut de Ci\`encies de l'Espai (ICE, CSIC) Campus UAB, Carrer de Can Magrans s/n E-08193 Cerdanyola del Vall\`es
Institut d'Estudis Espacial de Catalunya (IEEC) Edifici Nexus I, C/ Gran Capit\`a 2-4, despatx 201 E-08034 Barcelona, Spain
}

\author{L. Martin-Polo}
\affiliation{European Space Astronomy Centre, European Space Agency \\
Villanueva de la Ca\~{n}ada\\
28692 Madrid, Spain}

\author{J. Martino}
\affiliation{APC, Univ Paris Diderot, CNRS/IN2P3, CEA/lrfu, Obs de Paris, Sorbonne Paris Cit\'e, France}

\author{F. Martin-Porqueras}
\affiliation{European Space Astronomy Centre, European Space Agency \\
Villanueva de la Ca\~{n}ada\\
28692 Madrid, Spain}

\author{I. Mateos}
\affiliation{
Institut de Ci\`encies de l'Espai (ICE, CSIC) Campus UAB, Carrer de Can Magrans s/n E-08193 Cerdanyola del Vall\`es
Institut d'Estudis Espacial de Catalunya (IEEC) Edifici Nexus I, C/ Gran Capit\`a 2-4, despatx 201 E-08034 Barcelona, Spain
}

\author{P. W. McNamara}
\affiliation{European Space Technology Centre, European Space Agency\\ Keplerlaan 1\\ 2200 AG Noordwijk, The Netherlands} 
\author{J. Mendes}
\affiliation{European Space Operations Centre\\ European Space Agency\\ 64293 Darmstadt, Germany}

\author{L. Mendes}
\affiliation{European Space Astronomy Centre, European Space Agency \\
Villanueva de la Ca\~{n}ada\\
28692 Madrid, Spain}

\author{M. Nofrarias}
\affiliation{
Institut de Ci\`encies de l'Espai (ICE, CSIC) Campus UAB, Carrer de Can Magrans s/n E-08193 Cerdanyola del Vall\`es
Institut d'Estudis Espacial de Catalunya (IEEC) Edifici Nexus I, C/ Gran Capit\`a 2-4, despatx 201 E-08034 Barcelona, Spain
}

\author{S. Paczkowski}
\affiliation{Albert-Einstein-Institut, Max-Planck-Institut f\"ur Gravitationsphysik und Leibniz Universit\"at Hannover \\
Callinstra{\ss}e 38 \\
30167 Hannover, Germany}

\author{M. Perreur-Lloyd}                                                       
\affiliation{SUPA, Institute for Gravitational Research\\ School of Physics and Astronomy, University of Glasgow\\ Glasgow, G12 8QQ, UK}

\author{A. Petiteau}
\affiliation{APC, Univ Paris Diderot, CNRS/IN2P3, CEA/lrfu, Obs de Paris, Sorbonne Paris Cit\'e, France}

\author{P. Pivato}
\affiliation{Dipartimento di Fisica, Universit\`a di Trento and Trento Institute for Fundamental Physics and Application / INFN\\ 38123 Povo, Trento, Italy}

\author{E. Plagnol}
\affiliation{APC, Univ Paris Diderot, CNRS/IN2P3, CEA/lrfu, Obs de Paris, Sorbonne Paris Cit\'e, France}

\author{J. Ramos-Castro}
\affiliation{Department d'Enginyeria Electr\`onica, Universitat Polit\`ecnica de Catalunya\\  08034 Barcelona, Spain}

\author{J. Reiche}
\affiliation{Albert-Einstein-Institut, Max-Planck-Institut f\"ur Gravitationsphysik und Leibniz Universit\"at Hannover \\
Callinstra{\ss}e 38 \\
30167 Hannover, Germany} 

\author{D. I. Robertson}
\affiliation{SUPA, Institute for Gravitational Research\\ School of Physics and Astronomy, University of Glasgow\\ Glasgow, G12 8QQ, UK}

\author{F. Rivas}
\affiliation{
Institut de Ci\`encies de l'Espai (ICE, CSIC) Campus UAB, Carrer de Can Magrans s/n E-08193 Cerdanyola del Vall\`es
Institut d'Estudis Espacial de Catalunya (IEEC) Edifici Nexus I, C/ Gran Capit\`a 2-4, despatx 201 E-08034 Barcelona, Spain
}

\author{G. Russano}
\affiliation{Dipartimento di Fisica, Universit\`a di Trento and Trento Institute for Fundamental Physics and Application / INFN\\ 38123 Povo, Trento, Italy}


\author{J. Slutsky}
\affiliation{Gravitational Astrophysics Lab, NASA Goddard Space Flight Center\\ 8800 Greenbelt Road\\ Greenbelt, MD 20771 USA}

\author{C. F. Sopuerta}
\affiliation{
Institut de Ci\`encies de l'Espai (ICE, CSIC) Campus UAB, Carrer de Can Magrans s/n E-08193 Cerdanyola del Vall\`es
Institut d'Estudis Espacial de Catalunya (IEEC) Edifici Nexus I, C/ Gran Capit\`a 2-4, despatx 201 E-08034 Barcelona, Spain
}

\author{T. Sumner}
\affiliation{High Energy Physics Group, Physics Department, Imperial College London \\
Blackett Laboratory, Prince Consort Road \\
London, SW7 2BW, UK}

\author{D. Telloni}
\affiliation{INFN - Sezione di Firenze\\ via G. Sansone 1\\50019, Sesto Fiorentino, Firenze, Italy}
\affiliation{Osservatorio Astrofisico di Torino, INAF\\ Pino Torinese, Italy}

\author{D. Texier}
\affiliation{European Space Astronomy Centre, European Space Agency \\
Villanueva de la Ca\~{n}ada\\
28692 Madrid, Spain}

\author{J. I. Thorpe}
\affiliation{Gravitational Astrophysics Lab, NASA Goddard Space Flight Center\\ 8800 Greenbelt Road\\ Greenbelt, MD 20771 USA}

\author{D. Vetrugno}
\affiliation{Dipartimento di Fisica, Universit\`a di Trento and Trento Institute for Fundamental Physics and Application / INFN\\ 38123 Povo, Trento, Italy}

\author{M. Villani}
\affiliation{DISPEA, Universit\`a di Urbino ``Carlo Bo'' \\Via S. Chiara, 27 \\61029 Urbino, Italy}
\affiliation{INFN - Sezione di Firenze\\ via G. Sansone 1\\50019, Sesto Fiorentino, Firenze, Italy}

\author{S. Vitale}
\affiliation{Dipartimento di Fisica, Universit\`a di Trento and Trento Institute for Fundamental Physics and Application / INFN\\ 38123 Povo, Trento, Italy}

\author{G. Wanner}
\affiliation{Albert-Einstein-Institut, Max-Planck-Institut f\"ur Gravitationsphysik und Leibniz Universit\"at Hannover \\
Callinstra{\ss}e 38 \\
30167 Hannover, Germany}

\author{H. Ward}
\affiliation{SUPA, Institute for Gravitational Research\\ School of Physics and Astronomy, University of Glasgow\\ Glasgow, G12 8QQ, UK}

\author{P. Wass}
\affiliation{High Energy Physics Group, Physics Department, Imperial College London \\
Blackett Laboratory, Prince Consort Road \\
London, SW7 2BW, UK}

\author{W. J. Weber}
\affiliation{Dipartimento di Fisica, Universit\`a di Trento and Trento Institute for Fundamental Physics and Application / INFN\\ 38123 Povo, Trento, Italy}

\author{L. Wissel}
\affiliation{Albert-Einstein-Institut, Max-Planck-Institut f\"ur Gravitationsphysik und Leibniz Universit\"at Hannover \\
Callinstra{\ss}e 38 \\
30167 Hannover, Germany}

\author{A. Wittchen}
\affiliation{Albert-Einstein-Institut, Max-Planck-Institut f\"ur Gravitationsphysik und Leibniz Universit\"at Hannover \\
Callinstra{\ss}e 38 \\
30167 Hannover, Germany}

\author{P. Zweifel}
\affiliation{Institut f\"ur Geophysik, ETH Z\"urich\\ Sonneggstrasse 5\\ CH-8092, Z\"urich, Switzerland}





\begin{abstract}
Non-recurrent short term variations of the  galactic cosmic-ray (GCR) flux  above 70 MeV n$^{-1}$ were observed
between 2016  February 18 and 2017 July 3  aboard the European Space Agency LISA Pathfinder (LPF)  mission orbiting around the Lagrange point L1 at 1.5$\times$10$^6$ km from Earth.
The energy dependence of  three Forbush decreases (FDs) is studied and reported here. 
A comparison of these observations with others carried out in space down to the energy of a few tens of MeV n$^{-1}$ shows that
 the same GCR flux parameterization   
applies to events of different intensity during the main phase. 
FD observations in L1 with LPF and geomagnetic storm occurrence is also presented. 
Finally, the characteristics of GCR flux non-recurrent variations (peaks and depressions) of duration $<$ 2 days
and their association with interplanetary structures are investigated. It is found that, most likely, plasma compression regions 
between subsequent corotating high-speed streams cause peaks, while heliospheric current sheet crossing cause the majority of the depressions. 
\end{abstract}

\keywords{cosmic rays --- instrumentation: interferometers --- interplanetary medium --- Sun: heliosphere --- solar-terrestrial relations}



\section{Introduction} \label{sec:intro}

Galactic cosmic rays (GCRs) show an almost  isotropic spatial distribution in the inner heliosphere
and  consist of approximately  90\%  protons, 8\% helium nuclei, 1\%  heavy nuclei and 1\%  electrons  (percentages are in particle numbers to the total number).
The overall GCR energy integral flux at 1 a.u.  ranges approximately from 4000 particles m$^{-2}$   sr$^{-1}$ s$^{-1}$ at solar minimum through 1000  particles m$^{-2}$   sr$^{-1}$ s$^{-1}$ at solar maximum showing an eleven year   quasi-periodicity \citep[see for instance][]{pgs}.
During periods of negative solar polarity (when the global solar magnetic field lines enter the Sun North Pole) the flux of positively charged particles appears to be more modulated  up to a maximum  
of 40\% at 100 MeV n$^{-1}$ at solar minimum, with respect to epochs of positive solar polarity  \citep[when the global solar magnetic field lines exit the Sun North Pole,][]{pot13}.
The GCR flux modulation during  epochs of opposite solar polarities present a quasi-periodicity of twenty-two years  \citep[e. g.][and references therein]{lau14}.
In addition to these long-term GCR flux modulations, short-term variations  ($\leq$ 1 month) associated with the passage of large-scale 
interplanetary structures are also observed \citep[see for instance][]{richetal,richar,sabbah1,sabbah2,kudela,armano18a,pamelafd}.

LISA Pathfinder (LPF) was the key technology demonstrator mission of the European Space Agency (ESA)  Laser Interferometer Space Antenna (LISA), the first interferometer devoted  to gravitational wave detection in space in the
frequency interval 10$^{-4}$ Hz - 10$^{-1}$ Hz \citep{lisa}.      
The LPF spacecraft  orbited around the Lagrange point L1 at 1.5 million km from Earth in the Earth-Sun direction.
A high counting rate particle detector \citep[PD;][]{parde}, hosted aboard the
LPF mission \citep{lisapf1,lisapf2,armano,armano18b}  allowed for the measurement of the GCR integral proton and helium fluxes above 
70 MeV n$^{-1}$ from  2016 February 18 through 2017 July 3   during the descending phase of the present solar cycle N. 24 characterized by a positive polarity 
period of the Sun \citep{lisasymp,armano18a,armano18c}. 
The aim to place a PD aboard LPF was to measure the integral flux of particles of galactic and solar origin
energetic enough to penetrate the spacecraft and charge the test masses that constitute the heart of the interferometer.
Despite the PD was not meant for scientific use, it was tested on beam experiment \citep{mateos} and the minimum  energy 
of 70 MeV  n$^{-1}$ of ions crossing the detector  was measured with high accuracy.

This manuscript focuses on the characteristics of three Forbush decreases (FDs) and of  non-recurrent GCR flux short-term variations $<$ 2 days observed during the LPF mission lifetime.
FDs \citep{forb37,forb54,forb58,cane1} are sudden drops of the GCR flux intensity due the passage of interplanetary coronal mass ejections (ICMEs) and shocks.  These GCR non-recurrent variations  were primarily studied with the world wide neutron monitor (NM) network  since the 1950s \citep[see for instance][]{nm1,nm2}, although only cosmic-ray flux measurements gathered in space \citep{nm3}  allow for the study of the energy-dependence of the depressed GCR flux down to a few tens of MeV  without the use of models applied to Earth observations \citep{fluck,beer,uso11,uso17}.
 The LPF  2016 August 2 FD data \citep{armano18a} are compared  here to those  of the satellite experiment  PAMELA \citep{adriani,uso15,pamelafd}. 
The ratio of the depressed to pre-decrease  GCR fluxes  
during the main phase of the observed FDs is studied 
as a function of the energy. 

 FD, geomagnetic storm occurrence and the possibility of using FDs as precursor of geomagnetic activity was studied in the literature since early days  \citep[see for instance][]{nm3,badruddin,chauhan,kane10}. FD observations with LPF and contemporaneous geomagnetic activity  are illustrated here.

The low statistical errors characterizing the data provided by the PD aboard LPF allowed  also for the study of   GCR flux non-recurrent variations (depressions and peaks) shorter than 2 days.

This manuscript is organized as follows: in Section 2  the characteristics of the PD hosted aboard LPF  are described; in Section 3 the evolution of three FDs observed with LPF are compared to simultaneous  measurements of solar wind parameters carried out in L1 and to NM observations placed at different geographic latitudes; in Section 4  parameterizations of  proton and helium differential flux measurements gathered in space before and during the main phase of FDs are reported; in Section 5 a brief discussion on FDs and geomagnetic storm occurrence during LPF is presented and in Section 6 the association between  interplanetary structures and $<$2 day GCR flux non-recurrent variations is illustrated.

\section{The Particle Detector Aboard The LPF spacecraft}

The LPF spacecraft  was launched with a  Vega rocket from the Kourou base in  French Guiana on 2015 December 3.
The satellite reached its final 6-month orbit  around the  first Lagrangian point L1 
at the end of January  
2016.
The spacecraft elliptical orbit was inclined by about 45 degrees to the ecliptic. Minor and major axes of the orbit were approximately  0.5 million km and 0.8 million km, respectively. 

Two  nearly 2-kg cubic gold-platinum free-falling test masses played the role of mirrors  of the interferometer aboard LPF. 
Cosmic-rays with energies larger than 100 MeV n$^{-1}$ 
 penetrated approximately 13 g cm$^{-2}$ of spacecraft and instrument materials and charged the test masses. This process 
was expected to constitute one of the main  sources of noise for LISA-like space interferometers  in case of intense solar energetic particle events  \citep{shaul,wass}.
A PD  aboard LPF   allowed for $in$ $situ$  monitoring of  protons  and helium nuclei 
of GCRs and solar particles. A shielding copper  box of 6.4 mm thickness surrounded the silicon wafers in order to stop ions  with energies smaller than 70 MeV
n$^{-1}$. This conservative choice was made in order not to underestimate the overall incident particle flux  
charging the test-masses.   
The LPF PD was mounted  behind the spacecraft solar panels with its viewing axis along the Sun-Earth direction.
It consisted of two $\sim$ 300 $\mu$m thick silicon wafers 
of  1.40 x 1.05 cm$^2$ area,   placed in
a telescopic arrangement at a distance of 2 cm.  This detector allowed for the 
counting of particles traversing each of  the two silicon layers (single counts).  Single counts  were returned to the telemetry every 15 s. The energy deposits in the rear detector  of particles traversing both silicon wafers in less than 525 ns (coincidence mode) were stored  on the onboard computer in  histograms   of 1024 energy linear bins from 0 MeV to 5 MeV and returned to the telemetry every 600 seconds. The PD  geometrical factor   for particle  energies $>$ 100 MeV n$^{-1}$  was of 
  9 cm$^2$ sr for single counts and about  one tenth of this  value for  particles in coincidence mode. 
The maximum allowed detector  counting rate was 6500 counts s$^{-1}$ in the single count configuration.
In coincidence mode 5000 energy deposits per second   was the  saturation limit corresponding to an event   proton fluence of 
  10$^8$ protons cm$^{-2}$ at energies $>$ 100 MeV.

The spurious test-mass acceleration noise due  to the charging process  was  estimated  before the
mission launch with Monte Carlo simulations \citep{ara,gri05,wass05,gri15} on the basis of  GCR and solar energetic particle (SEP) flux predictions at the time the mission was supposed to be sent into orbit. 
The reliability of GCR flux predictions was positively tested with LPF data after mission end  \citep{armano18c} and with the Space Station AMS-02 magnetic spectrometer experiment  \citep{ams02} preliminary data above 400 MeV n$^{-1}$  presented at COSPAR 2018 (2018 July 14-22, Pasadena, USA) and expected to be reported in a forthcoming publication of the AMS collaboration. 
No SEP events occurred during the LPF mission, nevertheless 
 test-mass discharging was carried out  periodically with ultraviolet light beams illuminating the capacitor system surrounding the test-masses \citep{wass,armano18d} for acceleration noise 
control.

\section{Characteristics of Forbush decreases observed with LISA Pathfinder} \label{subsec:tables}

The LPF  15-s  proton and helium  single counts gathered between 2016 February 18 and 2017 July 3 were hourly-averaged in order to set  the statistical uncertainty of each data point  to 1\%.  
The  percentage change (PC) of these measurements calculated  with respect to their average value observed during  each Bartels rotation (BR) was visually inspected  over the LPF mission lifetime. This approach was adopted in order to limit the role of the solar modulation decrease during the years 2016-2017.  It is recalled here that the BR number represents the number of 27-day periods of the Sun since
 1832 February 8. 
The GCR flux variations  were then  compared to contemporaneous interplanetary magnetic field (IMF) and solar wind plasma parameters gathered by the ACE experiment \citep{stone} orbiting around the Lagrange point L1 \citep{nasacdaweb}.

 The passage of near-Earth ICMEs (reported in http://www.srl.calctech.edu/ACE/ASC/\\DATA/level3/icmetable2.htm)  
was associated with  three  FD observations carried out with the LPF PD on 2016 July 20, 2016 August 2 \citep [for this event see also][]{armano18a} and 2017 May 27 as it is shown in the left Figures 1-3. In these figures  the GCR flux variations are compared to the solar wind speed (V), to the IMF 
sunward x component in the GSE coordinate system with opposite sign (-B$_x$) observed to match the sector polarity and to the IMF intensity (B). The ICME passage is marked by dashed lines. The FD dated 2016 July 20 is associated with both solar wind speed and IMF increases  due to the ICME  propagating into a previous corotating high-speed solar wind stream (CHSS, V$>>$ 400 km s$^{-1}$). On 2016 August 2 and 2017 May 27 the GCR flux modulations appear correlated with the IMF intensity increase only. In all three cases the IMF intensity presented maximum values  of about 25 nT.  

\begin{figure}
\plottwo{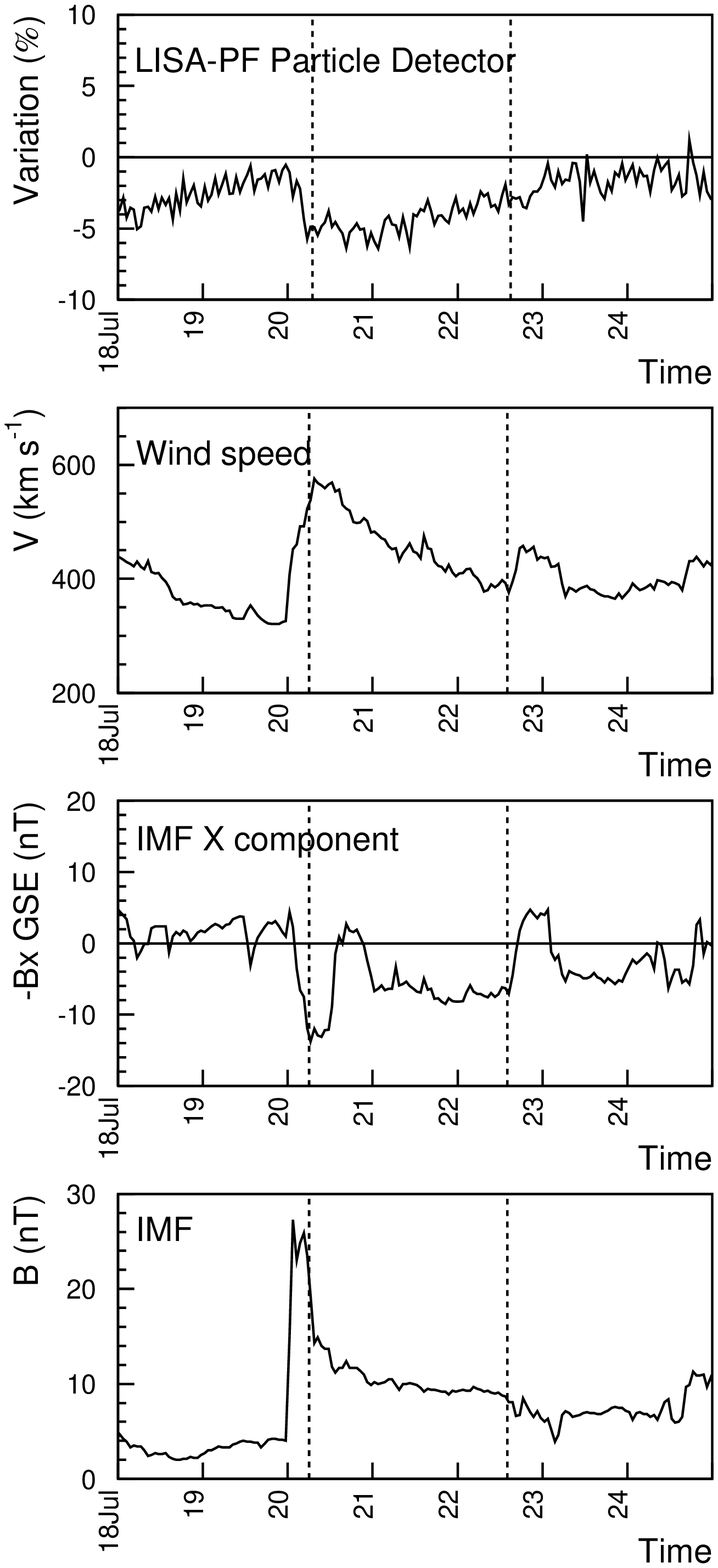}{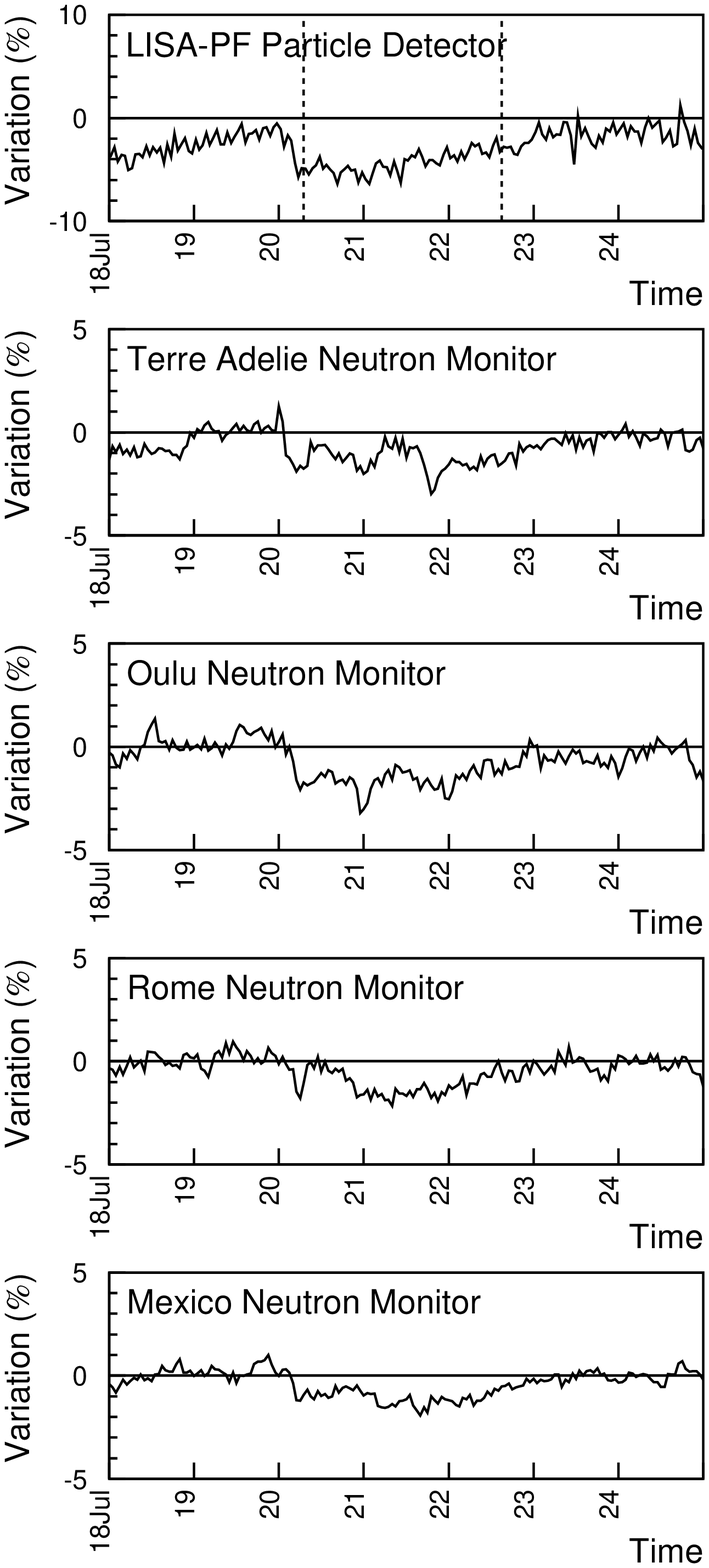}
\caption{Left: The LPF PD GCR hourly averaged counting rate PC between 2016 July 18  and 2016 July 24 are reported in the top panel. The solar wind speed (V) is shown in the second panel.  The Geocentric Solar Ecliptic (GSE) coordinate system sunward IMF  x-component with opposite sign (-B$_x$) and IMF intensity (B)  appear in the third and fourth panel, respectively. The IMF and solar wind parameter data were gathered from the ACE experiment \citep{nasacdaweb} in the Lagrange point L1.
The passage of a near-Earth ICME  is indicated by vertical dashed lines (http://www.srl.caltech.edu/ACE/ASC/DATA/level3/icmetable2.htm).
A FD is observed to begin on July 20. Right: Comparison of LPF  hourly-averaged GCR counting rate PC with
contemporaneous, analogous measurements of NMs  placed at various geographic latitudes \citep{nmonitors}.
Dashed lines in the top panel have the same meaning as those in the left figures.
\label{fig:f2}}
\end{figure}

\begin{figure}
\plottwo{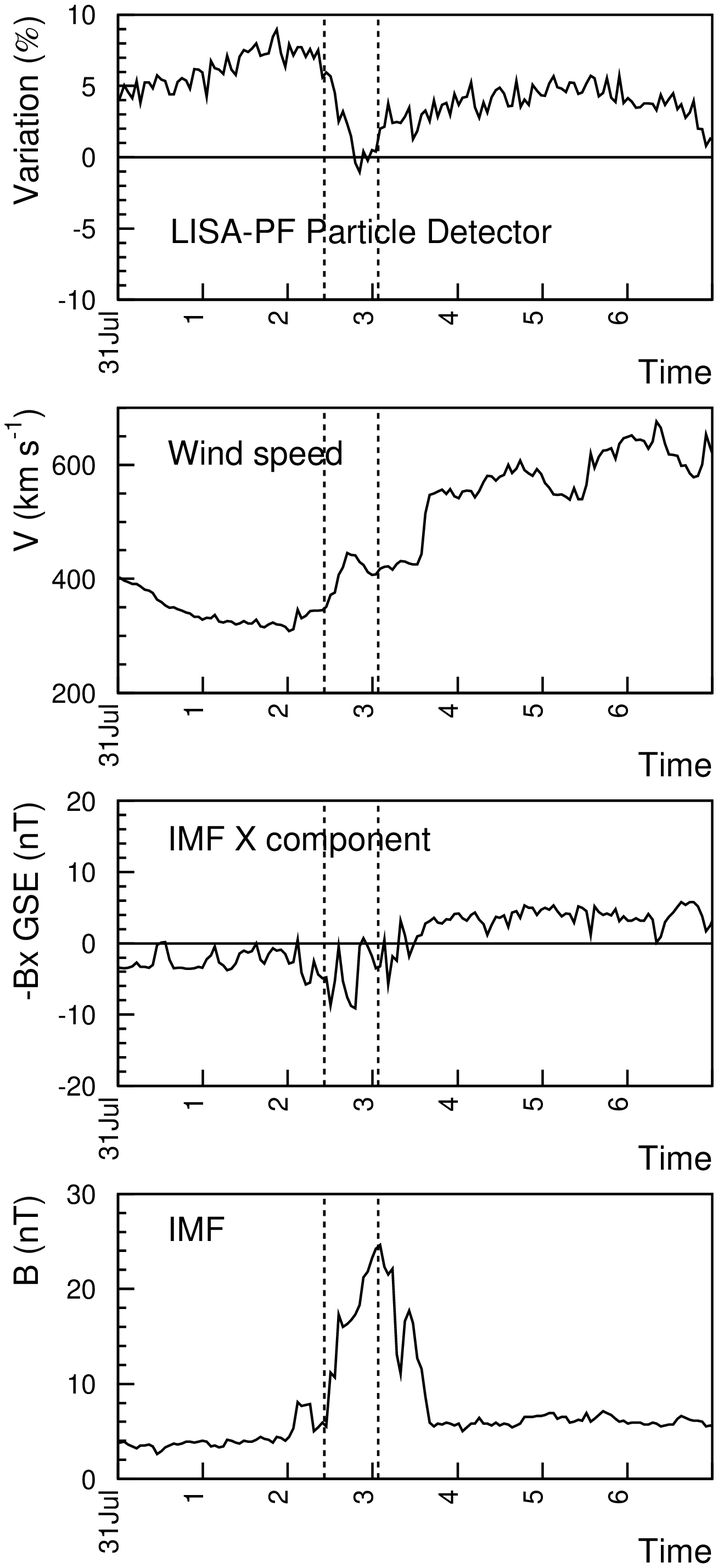}{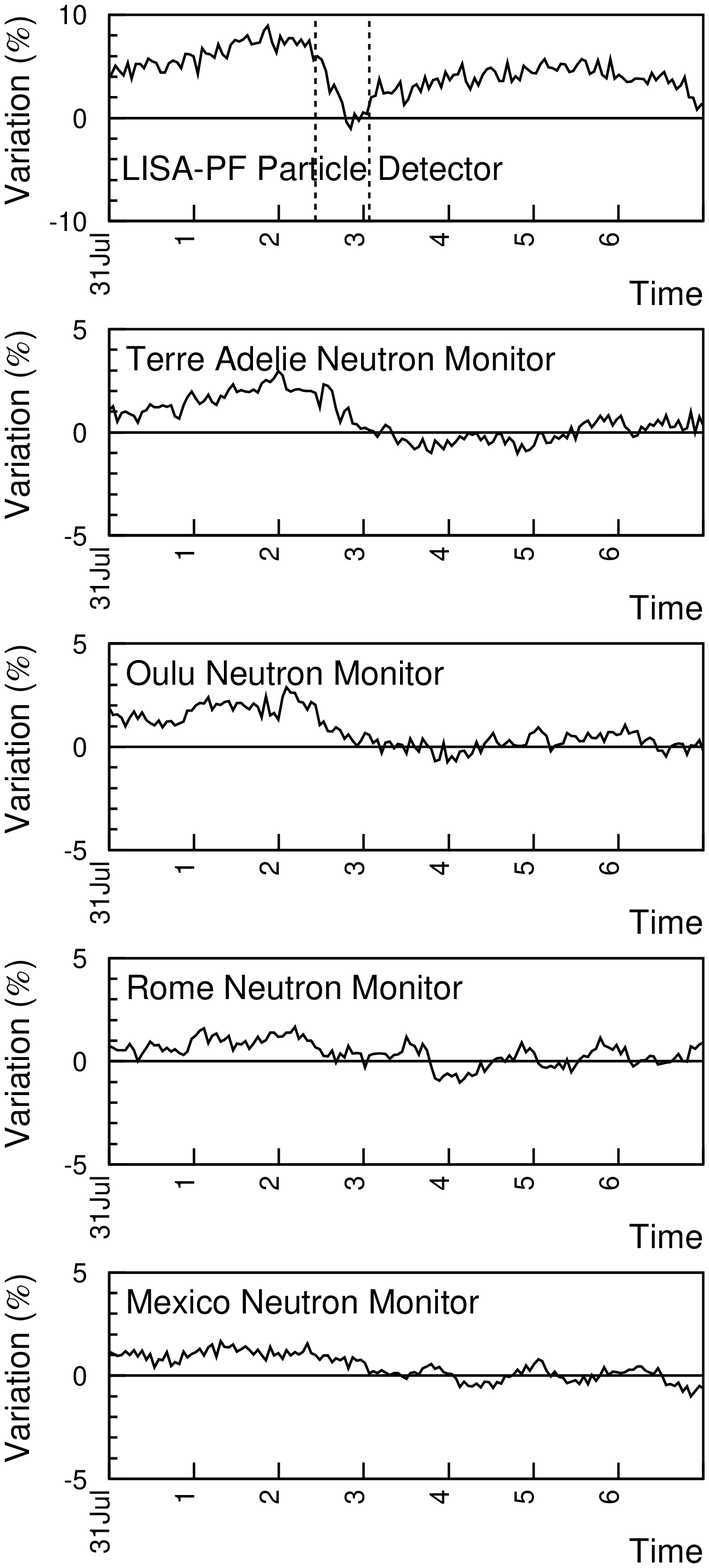}
\caption{Same as Figure 1 for the period 2016 July 31 - 2016 August 6.\label{fig:f2}}
\end{figure}

\begin{figure}
\plottwo{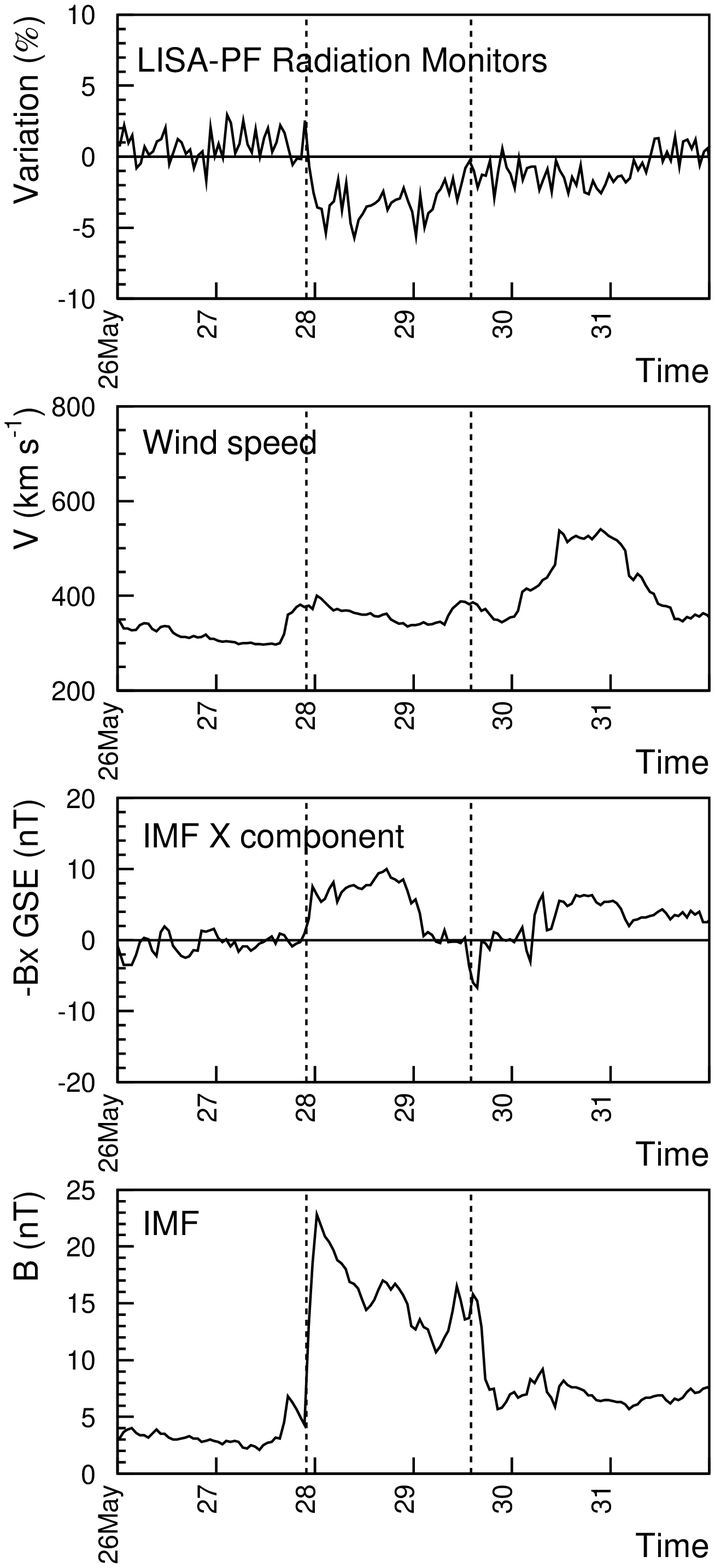}{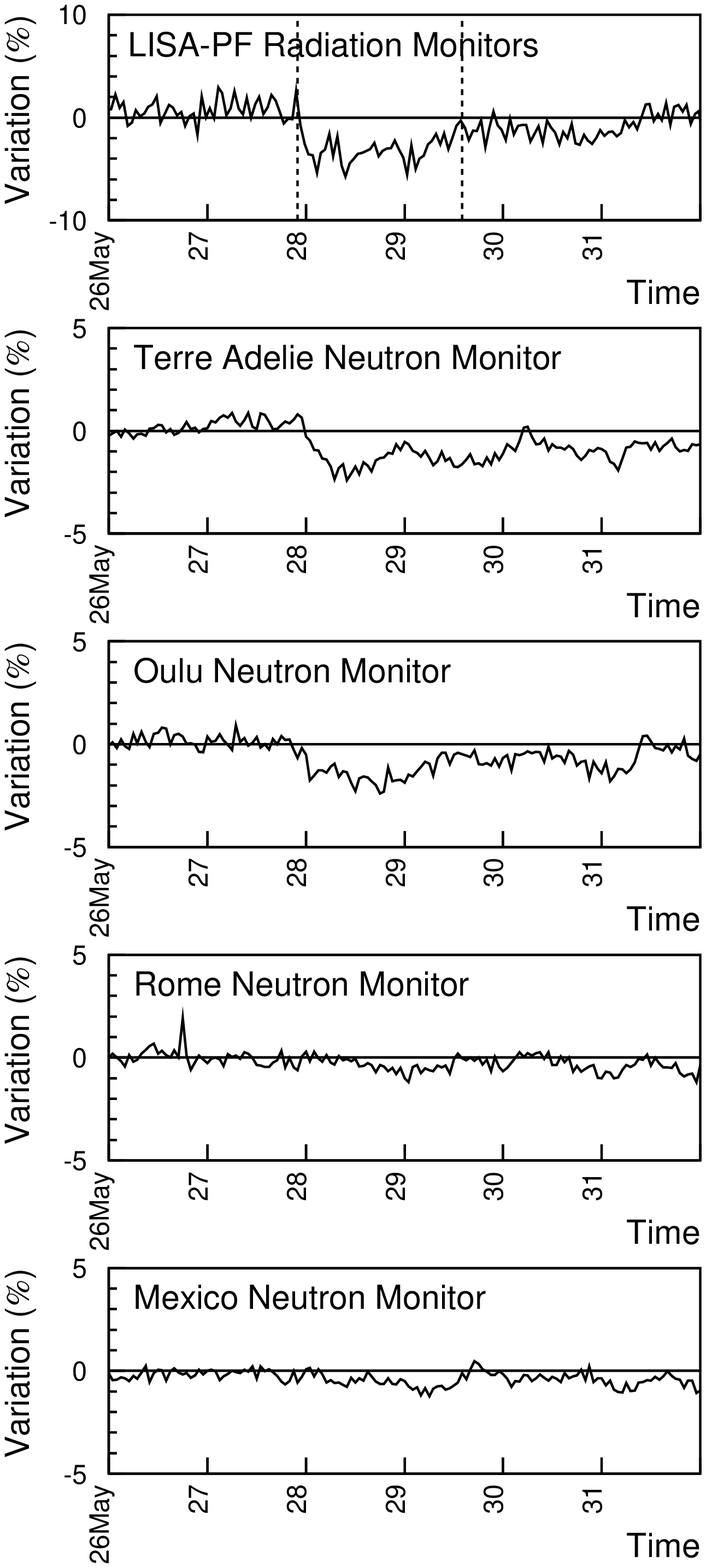}
\caption{Same as Figure 1 for the period 2017 May 26 - 2017 May 31.\label{fig:f2}}
\end{figure}

In order to study the energy dependence of the three FDs observed with LPF,   the PC  of the  integral proton and helium fluxes measured with the PD above 70 MeV n$^{-1}$ were compared  
to contemporaneous hourly-averaged PC of observations
gathered with NMs located at different geographic latitudes in the right Figures 1-3 (www.nmdb.eu, a similar attempt for NMs only was carried out in \citet[][]{badruddin}). The GCR flux PC observed aboard LPF or with NMs was calculated by using as baseline (PC=0\%)  the  average values of counts measured by PD or NMs  during the  BR to which the studied period of time belongs. 
The Terre Adelie,  Oulu, Rome and Mexico NM stations are characterized by geomagnetic cut-off rigidities of 0 GV, 0.8 GV, 6.3 GV and 8.2 GV, respectively. The shielding effect of the atmosphere and  the geomagnetic cut-off do not allow NMs of providing direct measurements of  GCR energy spectra at low energies. Conversely,  the PC of NM counting rate measured on Earth is approximately the same of the GCR integral flux incident at the top of the atmosphere above  $effective$ $energies$. Effective energies range from 11-12 GeV for polar stations through 20 GeV for equatorial stations \citep[see for details][and references therein]{gil3}. In Table 1 are reported the PC  of the GCR integral flux observed with LPF  and with the NMs listed above, at the maximum of each FD.  Time of the onset and of the maximum  of each FD aboard LPF are also indicated. The onset  was set as the first time bin after which the GCR flux presented a continuous decrease trend, within statistical uncertainty,  for at least six hours. The time when the GCR integral flux reached its minimum value during each FD was estimated with a   best line fit through the data points.  
The LPF  proton-dominated (resulting from proton and helium measurements) integral flux maximum decreases above 70 MeV were observed to vary from about 5\% to 9\% during the three events.  The different GCR flux decrease observed with LPF in response to similar IMF intensity increases  is most likely due to the  passages of interplanetary structures that depressed the GCR flux before the transit of the ICMEs. During the 2016 August 2 event only, the pre-decrease GCR flux appeared at its maximum value during the BR 2496 before the passage of the ICME that generated the FD (see Figure 7 in \citet[][]{armano18a}).
  NM data show  PCs above effective energies ranging between 1\% and 3\%. 
Both GCR flux main and recovery phases are observed in all considered NM measurements during the 2016 July 20 FD. This is not the case for the other two events that can be clearly detected in polar NM measurements only. The
 energy dependence of GCR flux depressions during FDs was also discussed, for instance, in \citet[][]{uso08,grim11,badruddin}. 

\startlongtable
\begin{deluxetable}{ccccccc}
\tablecaption{Energy dependence of the GCR integral flux PC at the maximum of the  three FDs observed aboard LPF above 70 MeV n$^{-1}$ and with NMs above effective energies: 11 GeV for  polar stations; 12 GeV for Oulu NM;  17 GeV for Rome NM and 20 GeV for Mexico NM \label{tab:table}}
\tablehead{
  \colhead{LPF FD onset} &\colhead{LPF  FD maximum} &\colhead{PC} & \colhead{PC}& \colhead{PC}& \colhead{PC}&\colhead{PC}\\
\colhead{Time} & \colhead{Time} &\colhead{$>$ 70 MeV} & \colhead{$>$ 11 GeV} & \colhead{$>$ 12 GeV} & \colhead{$>$ 17 GeV}&  \colhead{$>$ 20 GeV}
}
\startdata
     2016 July 20 07.00 UT &  July 21 01.00 UT& 5.5\%  & 2\% & 2\% & 2\% & 1\%\\
     2016 August 2 12.00 UT & August 2 22.40 UT& 9\%   & 3\% & 2\% & 2\% & 1\%\\
     2017 May 27 18.00 UT & May 28 10.45 UT&  7\% & 3.5\% & 2.5\% & 1\% & 1\%\\
\enddata
\end{deluxetable}

\begin{figure}
\begin{center}
  \includegraphics[width=0.7\textwidth]{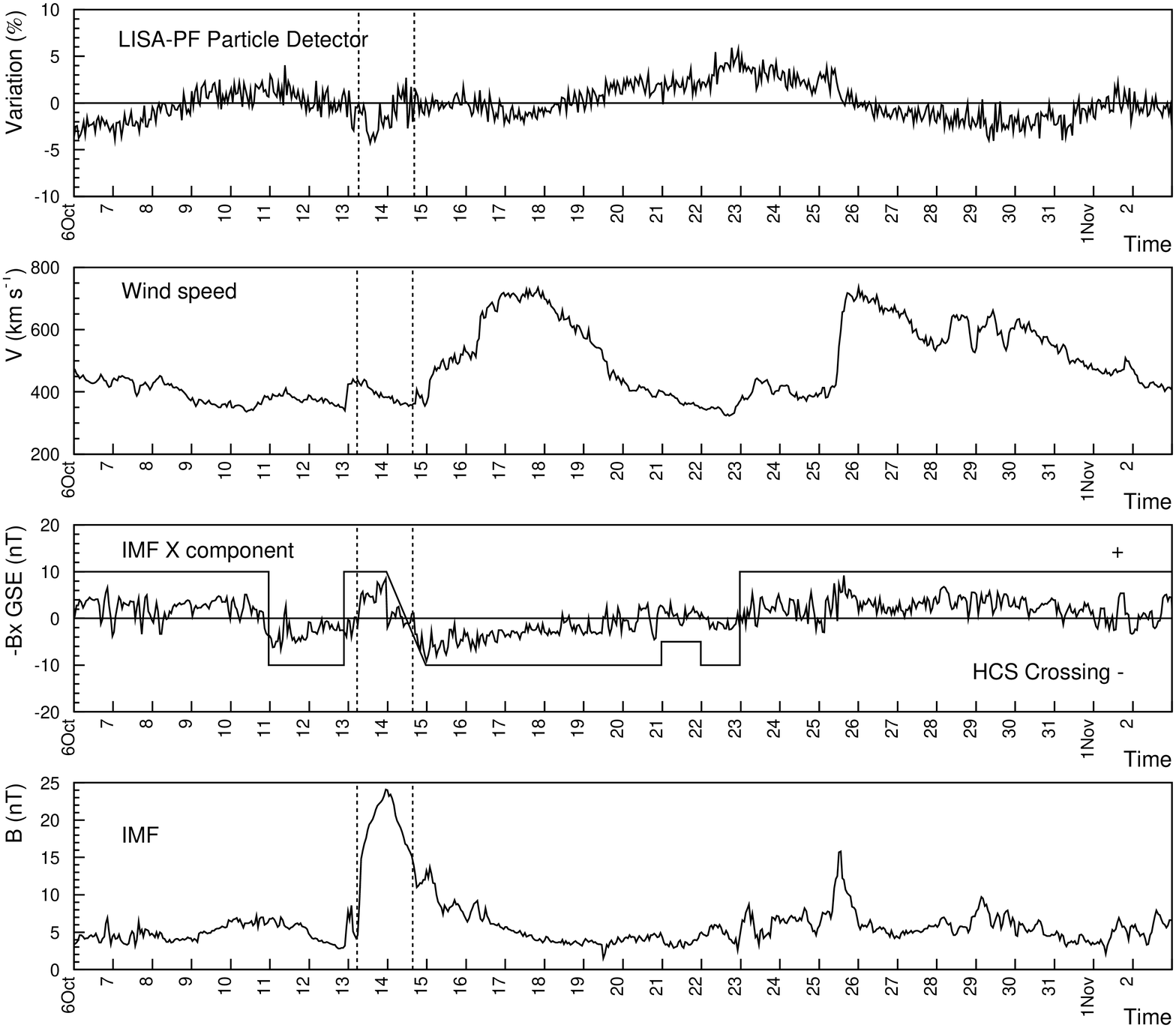}
  \caption{From top to bottom panels: GCR flux variations aboard LPF; solar wind speed (V), IMF negative component (-B$_x$) in GSE coordinate system and IMF intensity (B) during the  BR 2499 (2016 October 6 - 2016 November 1). In panel 3 the continuous line indicates HCSC and sector daily polarity (positive and  negative polarities were set to +10 and -10 arbitrarily in the plot). Undefined polarities were equally arbitrarily  set to +5 and -5 (sector polarities are reported in \url{http://omniweb.sci.gsfc.nasa.gov./html/polarity/polarity\_tab.html}). The passage of a near-Earth ICME  is indicated by vertical dashed lines (http://www.srl.caltech.edu/ACE/ASC/DATA/level3/icmetable2.htm).}
  \label{figure1}
\end{center}
 \end{figure}

The transit of other three near-Earth ICMEs on 2016 March 5, 2016 April 14 and 2016 October 13  resulted in   GCR flux decreases  at the limit of the statistical significance  (1-2\%) on LPF, the GCR  flux being already  reduced by the transit of previous interplanetary structures  and heliospheric current sheet crossing (HCSC; see  Figure 6 in \citet{armano18a} and Figure 4).

\section{Parameterization of GCR energy spectra during FDs}

GCR flux measurements gathered  in space  are considered in this Section to investigate if the same parameterization could be used to replicate the trend of the GCR flux  PC during the main phase of FDs of different intensity. 
The  LPF GCR observations gathered before and at the maximum (22.40 UT) of the 2016 August 2  FD  are compared to those of the satellite
PAMELA experiment,  which measured  both proton and helium  differential fluxes before and during the main phase of the  FD dated  2006 December 14 between 16.50 UT and 22.35 UT \citep[see for details][]{adriani,uso15,pamelafd}. The PAMELA data can be found in https://tools.ssdc.asi.it/CosmicRays/. In Figure 5 vertical solid lines delimit the interval of time during which PAMELA observed the FD. The pre-decrease and depressed proton energy spectra   
observed by PAMELA in November 2006 and on  2006 December 14, respectively, 
are shown in Figure 6. The PAMELA data are reported in https://tools.ssdc.asi.it/CosmicRays/. In the same figure,  pre-decrease and depressed proton dominated energy differential fluxes  for the  FD dated 2016 August 2  are also reported.
\begin{figure}
  \begin{center}
  \includegraphics[width=0.7\textwidth]{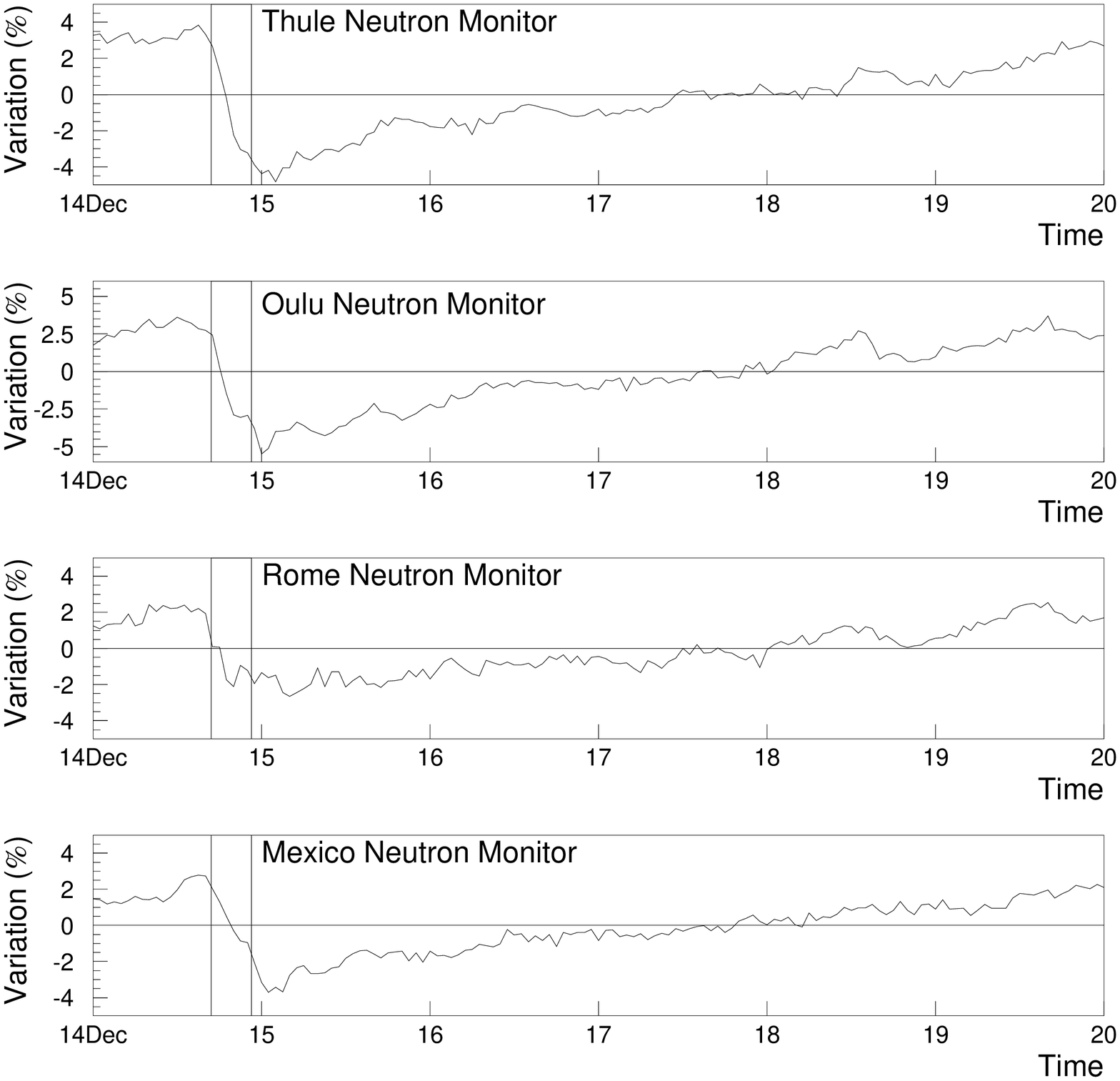}
  \caption{NM measurements at various geographic latitudes between 2006 December 14 and 2006 December 19. A FD on December 14
 was also observed in space
by the satellite experiment PAMELA above 70 MeV n$^{-1}$ from 16.50 UT through 22.35 UT on 2006 December 14 (vertical continuous lines).}
  \label{figure1}
 \end{center}
\end{figure}
\begin{figure}
  \begin{center}
  \includegraphics[width=0.7\textwidth]{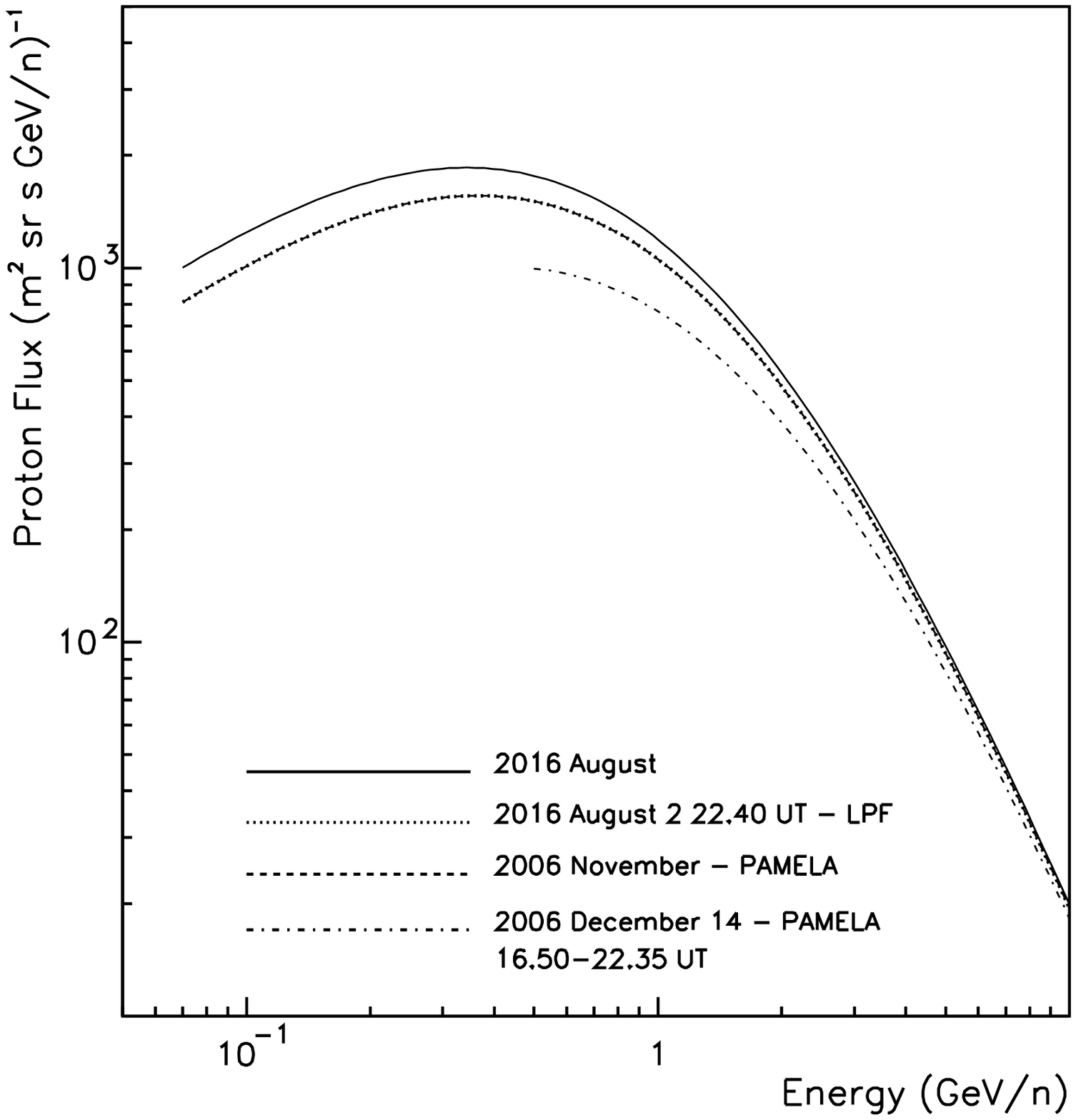}
  \caption{Proton energy spectra measured by the PAMELA experiment before (dashed line) and during  the FD dated 2006 December 14 (dot-dashed line).
 The proton-dominated  observations carried out with LPF and NMs on
2016 August 2 at the maximum of the FD (22.40 UT) are also shown (dotted line). The 2016 August pre-decrease proton flux is represented by a continuous line. The depressed proton spectrum observed during the August 2 FD with LPF superpose to the pre-decrease proton flux measured by PAMELA in November 2006.}
  \label{figure1}
 \end{center}
\end{figure}
The PAMELA data gathered during the main phase of the FD are not shown below 500 MeV
since the proton flux included particles of both galactic and solar origin. It is pointed out that the PAMELA pre-decrease flux measurements were 
considered those gathered in 2006 November   since the solar modulation during the months of 2006  November and 2006 December  was very similar \citep{oulu}.
PAMELA helium  data for the same FD appear in Fig. 7. 
\begin{figure}
  \begin{center}
  \includegraphics[width=0.7\textwidth]{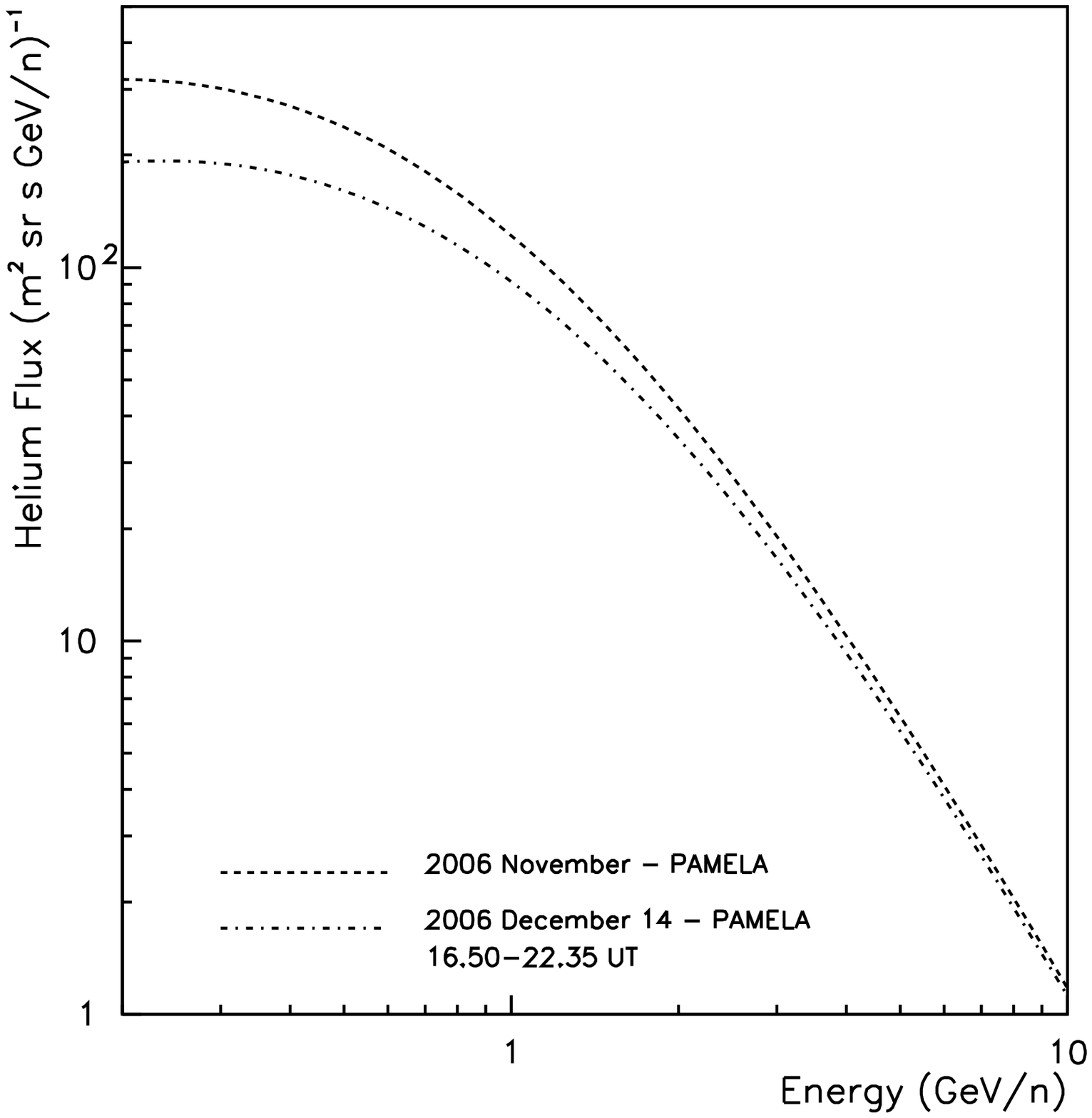}
  \caption{Helium energy spectra measured by 
the  PAMELA experiment before and during the FD dated 2006 December 14.}
  \label{figure1}
 \end{center}
\end{figure}

 The  energy spectra of cosmic rays observed during the main phase of the FDs (F$_{FD}$(E)) considered in this Section and corresponding pre-decrease energy spectra (F(E)) are parameterized as indicated in eqs. 1 and 2  \citep{pgs}, respectively: 

\begin{equation}
F_{FD}(E)= A\ (E+b')^{-\alpha}\ E^{\beta}  \ \ \ {\rm particles\ (m^2\ sr\ s\ GeV n^{-1})^{-1}},
\label{equation2}
\end{equation}

\begin{equation}
F(E)= A\ (E+b)^{-\alpha}\ E^{\beta}  \ \ \ {\rm particles\ (m^2\ sr\ s\ GeV\ n^{-1})^{-1}}
\label{equation2}
\end{equation}

with $b'$ $>$ $b$. The parameters $\alpha$ and $\beta$ remain unchanged since these parameters modulate the GCR flux above 10 GeV where pre-decrease and depressed fluxes present approximately the same slope as it can be observed in Figures 6 and 7. 
The parameterizations reported in eqs. 1 and 2 are found  
to reproduce the GCR energy spectra trend in the inner heliosphere in the energy range  of observations between a few tens of MeV n$^{-1}$ up to 
hundreds of GeV n$^{-1}$ in agreement with the Gleeson and Axford model \citep{gle68} within  experimental errors of data. 
These parameterizations   are adopted in this work instead of using the model by Gleeson and Axford 
during FDs \citep[see for instance][]{uso15}, since in this last model the modulation of GCR energy spectra is correlated with the solar modulation parameter that follows the long-term quasi-periodicity of the solar activity. 
The solar modulation parameter is kept  here constant during each BR and  it is preferred  to 
 increase the parameter b (in eq. 2) to b' (in eq. 1) to reproduce the observed GCR flux trend during a FD to decouple the effects of long and short-term GCR flux variations. 




The parameters $A$, $b$, $b'$, $\alpha$, $\beta$  estimated for each data set are indicated in Table 2. The $\chi$$^2$ and number of  degrees of freedom  for  each parameterization of the PAMELA data (available in https://tools.ssdc.asi.it/CosmicRays) are also  reported in Table 2.
 For the 2016 August 2 LPF FD  
the pre-decrease proton differential flux 
  above 70 MeV was estimated on the basis of the Gleeson and Axford model  by assuming a solar modulation parameter of 438 MV  
\citep{oulu} for the 2016 August month and the 
interstellar proton spectrum by \citet{burger}. The differential flux thus obtained was then parameterized as indicated in eq. 2 and integrated  above 70 MeV and above the effective energies of polar, Oulu, Rome and Mexico NM stations. The integral flux values were then reduced at 70 MeV and at effective energies as observed by LPF at 22.40 UT of 2016 August 2 and by NMs between 22.00 and 23.00 UT of the same day. 
 Finally, the  differential flux  at the maximum of the FD  was  estimated by  increasing the parameter $b$ of the pre-decrease differential flux (third raw in Table 2 and eq. 2) to $b'$ (fourth raw in Table 2 and eq. 1) until obtaining an agreement to better than 1\% between the modulated  integral flux  and  integral flux measurements carried out with LPF and NMs. No $\chi$$^2$  was calculated for LPF since no differential flux measurements are available for our experiment. In \citet{armano18a} the same approach presented here was adopted by using, however, the \citet{bess} interstellar proton spectrum inferred from the BESS experiment data gathered during both positive and negative polarity periods of the Sun. The solar modulation parameter values obtained with the BESS data differ from those reported in \citet{oulu} obtained with the \citet{burger} interstellar spectra only by a few tens of MV, considered to lie within the uncertainty of the method. After the publication of the AMS experiment data gathered  during the month of August 2016, it will be possible to set the uncertainties on the outcomes of the present work.

By defining R(E) the ratio of  the GCR flux interpolations that appear in eqs. 1 and 2, respectively:

\begin{equation}
 R(E)= \frac{F_{FD}(E)}{F(E)},
\label{equation1}
\end{equation}


it is found:

\begin{equation}
 R(E)=  \left(\frac{E+b'}{E+b}\right)^{-\alpha}.
\label{equation1}
\end{equation}

\begin{deluxetable}{cccccccc}
\tablecaption{Parameterizations of proton (p) and helium (He) energy spectra measured by the indicated experiments before and during  Forbush decreases (see eqs. 1 and 2). The $\chi$$^2$ and number of degrees of freedom ( $ndof$) estimated for each set of experimental data are indicated.\label{tabparam}}
\tablehead{
\colhead{} & \colhead{$A$} & \colhead{$b$} & \colhead{$b'$} & \colhead{$\alpha$}& \colhead{$\beta$} &{$\chi$$^2$}& \colhead{$ndof$}
}
\startdata
p (PAMELA experiment - 2006 November)  & 18000 & 1.17 & & 3.66 & 0.87 & 2279.1 & 71\\
p (FD - 2006 December 14 16.50 UT - 22.35 UT)  & 18000 &  & 1.37 & 3.66 & 0.87 & 4948.7 & 71\\
p (LPF - 2016 August)  & 18000 & 1.10 & & 3.66 & 0.87 & - & - \\
p (2006 August 2 22.40 UT)  & 18000 &  & 1.17 & 3.66 & 0.87 & - & - \\
He (PAMELA experiment - 2006 November) & 850  & 0.75  & &3.47 & 0.72 &16.38 & 18\\
He (FD - 2006 December 14 16.50 UT - 22.35 UT) & 850  & &0.90  & 3.47 & 0.72&10.98& 18\\
\enddata
\end{deluxetable}
R(E) estimated for  LPF and PAMELA proton measurement interpolations  are shown in Figure 8 while in Figure 9, R(E) was calculated for the PAMELA  helium observation interpolations. 
The simple relationship in eq. 3 allows for a quick, even though approximate, estimate of the GCR energy differential flux during the main phase of a FD when integral flux measurements during the event evolution and the differential flux before the occurrence of the same  are known.

The $b/b'$ ratios of the  parameters estimated with the GCR flux parameterizations before and during the main phase of each FD studied in this Section (two data points were considered for the FD dated 2006 December 14 since PAMELA measured both proton and helium fluxes) appear correlated with the GCR flux percentage attenuation (PA)  as it is shown in  Figure 10 (solid dots). PA is defined as follows:

\begin{equation}
 PA=  \frac{\int_{}^{} F_{FD}(E) dE}{\int_{}^{} F(E) dE},
\label{equation1}
\end{equation}

where integrals are calculated in the energy range of data availability during each event.

In Figure 11 
the continuous  line indicates the best fit through the data points. 
If additional observations gathered in space will confirm the reliability of this simple empirical relationship, it will be possible to set a statistical significance for the same. 

\begin{figure}
  \begin{center}
  \includegraphics[width=0.7\textwidth]{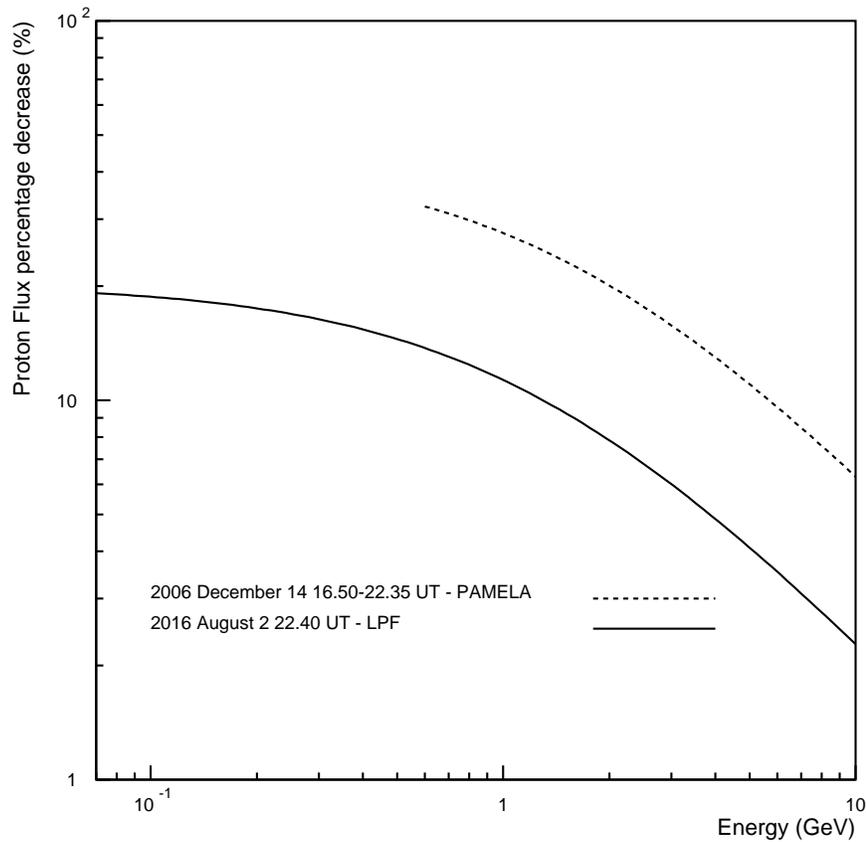}
  \caption{Parameterization of the proton flux percentage decrease observed by PAMELA and LPF during the main phase of the FDs dated on 2006 December 14 and 2016 August 2, respectively.}
  \label{figure1}
 \end{center}
\end{figure}

\begin{figure}
  \begin{center}
  \includegraphics[width=0.7\textwidth]{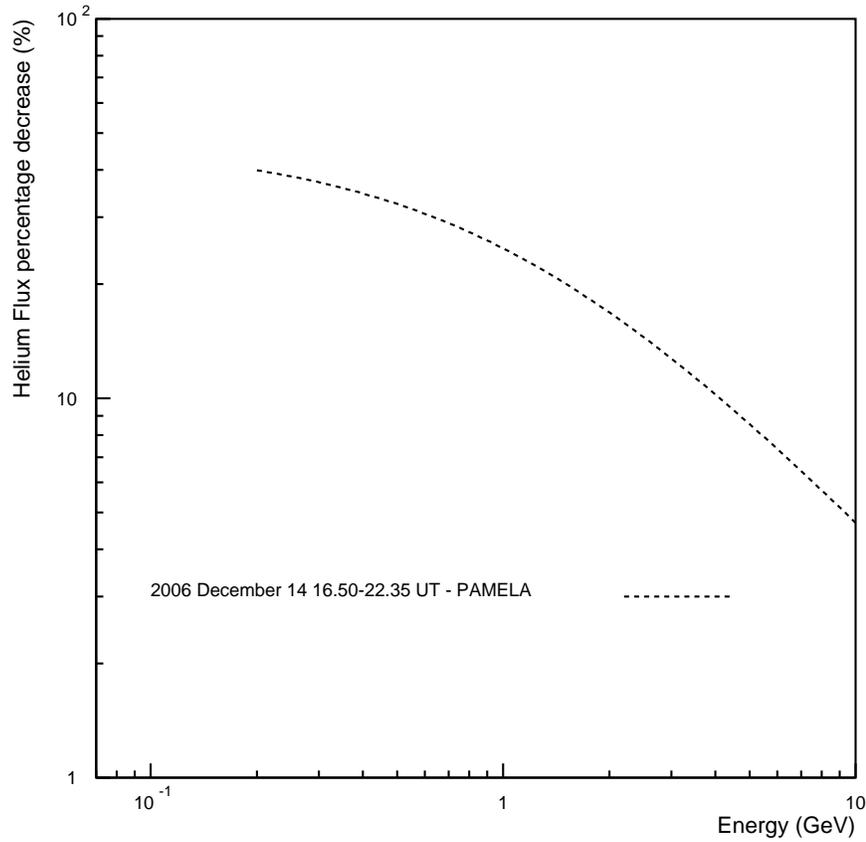}
  \caption{Parameterization of the percentage decrease of  the helium flux observed by the 
PAMELA experiment during the FD dated on 2006 December 14.}
  \label{figure1}
 \end{center}
\end{figure}

\begin{figure}
  \begin{center}
  \includegraphics[width=0.7\textwidth]{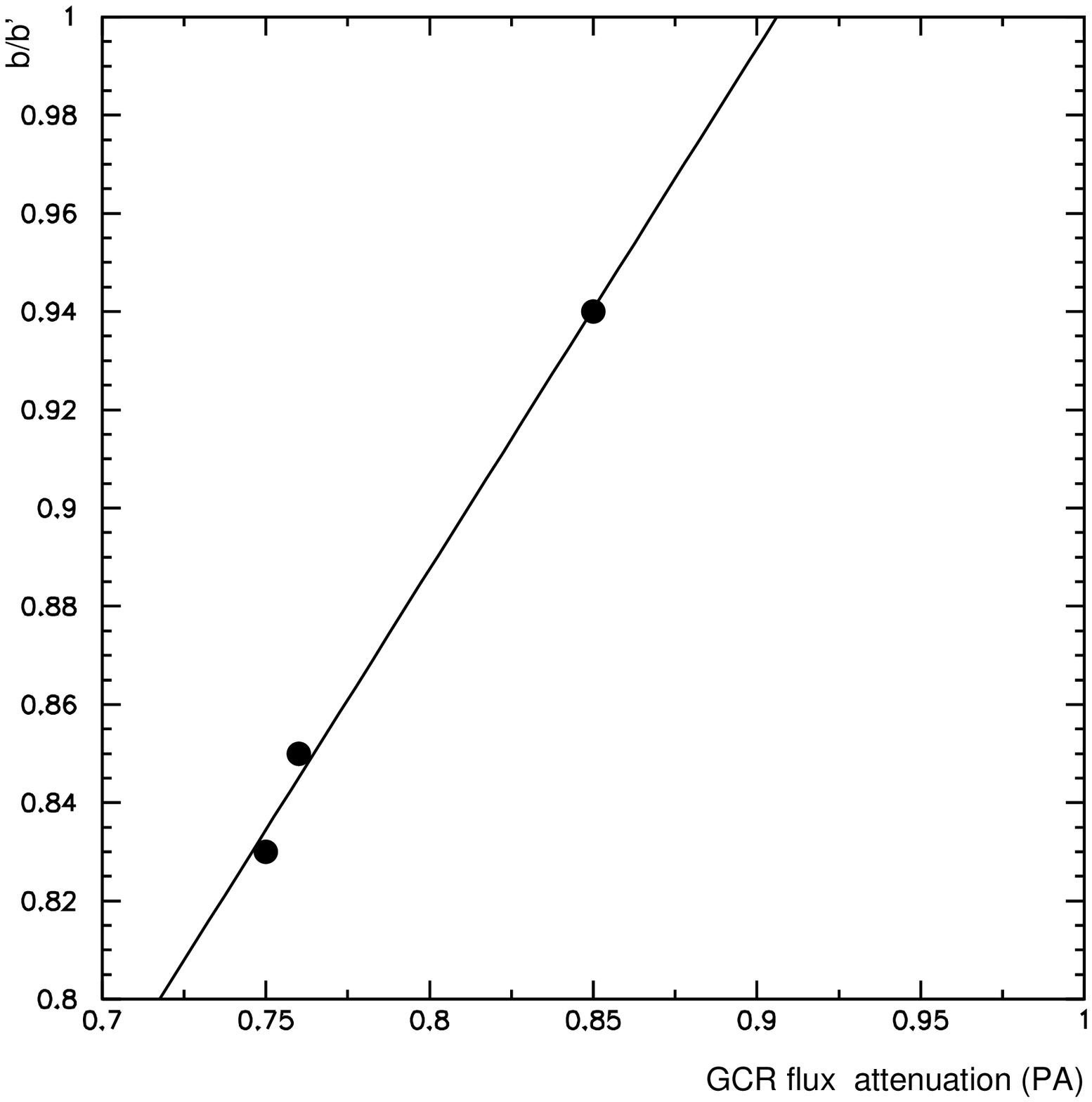}
  \caption{Parameterization  of the $b/b'$  ratio inferred from eqs. 1 and 2 for the FDs studied in this Section versus the GCR flux percentage attenuation (PA). The continuous line indicates the best-fit of the data points: $b/b'$=1.061 PA+0.0387. 
}
  \label{figure1}
 \end{center}
\end{figure}



\section{FD observations in the Lagrange point L1 and geomagnetic storm occurrence}

Fifteen  near-Earth ICMEs were observed  (http://www.srl.caltech.edu/ACE/ASC/DATA/level3\\/icmetable2.htm) during the time  the LPF spacecraft was orbiting around the Lagrange point L1. Eight of these ICMEs presented magnetic clouds. As it was anticipated in Section 3, the GCR integral flux measurements aboard LPF  presented  depressions at the time of the passage of  six of these  ICMEs (2016 March 5; 2016 April 14; 2016 July 20; 2016 August 2, 2016 October 13 and 2017 May 27)  but in three cases only (on 2016 July 20, 2016 August 2 and 2017 May 27)  FDs were observed.
During  the main phases of these three FDs the GCR flux decreases  appeared  correlated with the increase of the IMF intensity up to about 25 nT associated with the contemporaneous transit of ICMEs \citep[see also][]{benella} while the solar wind speed remained below 400 km s$^{-1}$ except at the onset of the 2016 July 20 event. Conversely, during the GCR flux depressions observed on 2016 March 5 (Figure 6 in Armano et al.; 2018a) and 2016 April 14 (Figure 11) the effects of  ICME passages (from 19.00 UT on March 5 through 15.00 UT on March 6 and 09.00 UT on April 14 through 04.00 UT on April 15, respectively) were mainly concealed by the action of concomitant transits of several CHSS. On 2016 October 13 the role of  a near-Earth ICME passage  (from 2016 October 13 at 6.00 UT through 2016 October 14 at 16.00 UT; dashed lines in Figure 4) and  increase of the IMF intensity $>$ 20 nT in modulating the GCR flux could not be established since the GCR flux presented a continuous decreasing trend well before the passage of the ICME due to a previous transit of high-speed solar wind streams  and  HCSC on October 11, October 13 and October 14.

Geomagnetic storms are disturbances of the Earth's magnetosphere classified on the basis of their intensity by changes in the Dst (disturbance storm time) geomagnetic index representing the average change of the horizontal component of the Earth's magnetic field at the magnetic equator \citep{gonzales}. 
Geomagnetic storms are defined  weak when the Dst ranges between -30 nT and -50 nT; moderate when the Dst varies between -50 nT and -100 nT and strong when the Dst is smaller than -100 nT. Moderate geomagnetic storms, more frequent than strong ones, affect communications while the most intense  ones may induce  severe damages in critical infrastructures on Earth.  Near-Sun coronal mass ejection and near-Earth solar wind parameters are used to forecast geomagnetic storms \citep{kim}. 
The geomagnetic index Dst  reached a value smaller than -100 nT  twice during the period LPF collected data (2016 February 18 - 2017 July 3):  
on 2016 October 13 at 17.30 UT (-104 nT) and  on 2017 May 28 at 07.30 UT (-122 nT). In Figure 12 it is shown that no FD can be observed beyond statistical fluctuations with LPF 
and NMs.
Conversely, the passage of a near-Earth ICME was  at the origin of both the FD observed on LPF and NMs on 2017 May 27-28 (Figure 3) and the geomagnetic 
storm  occurred on 2017 May 28.
The 2016 August 2  FD onset occurred at 12.00 UT aboard LPF, about ten  hours before a weak   geomagnetic storm (Dst $\simeq$ -50 nT) that started at 22.00 UT when the FD reached its maximum in the Lagrange point L1 \citep[see Figure 7 in][]{armano18a}. For this event the geomagnetic storm and the maximum of the FD occurred at the same time even though this is not a general result  \citep[see][for instance]{kane10}. 
Geomagnetic storms are caused by fast solar wind streams and large negative 
values of the B$_z$ component of the IMF reconnecting with the Earth magnetic field while FDs are caused by large increases of the IMF intensities.  
FD  observations with LPF in the Lagrange point L1 and geomagnetic storm occurrence  are summarized in Table 3 along with 
 maximum values of the observed IMF intensity and minimum values of the B$_z$ component during each FD.  It can be concluded that FDs, when observed,  can be used to forecast geomagnetic storms only when  the  z-component of the IMF presents values $<$ -20 nT \citep[see also][]{dremukhina}.
 

\begin{deluxetable}{cccccc}
\tablecaption{FD observations and  geomagnetic storm occurrence  during the LPF mission. \label{tabparam}}                                                                              
\tablehead{                                                                                                                                                          
\colhead{Date} &  \colhead{FD} & \colhead{Geomagnetic storm} & \colhead{Maximum B} & \colhead{Minimum Bz}& \colhead{Dst}\\
\colhead{} & \colhead{Yes/No} &\colhead{Yes/No} & \colhead{nT} & \colhead{nT}& \colhead{nT}}
\startdata                                                                                                                                                           
2016 July 20  &Yes & No & 25  & -8.9 & $>$ -50\\                                                                                                                       
2016 August 2 & Yes& No & 24 & -9.5 & $\simeq$ -50\\                                                                                                                        
2016 October 13 & No & Yes & 24 & -19 & -102\\                                                                                           
2017 May 27   & Yes&Yes & 23 &  -21& -122\\                                                                                            
\enddata                                                                                                                                                             
\end{deluxetable}

\begin{figure}
  \begin{center}
  \includegraphics[width=0.7\textwidth]{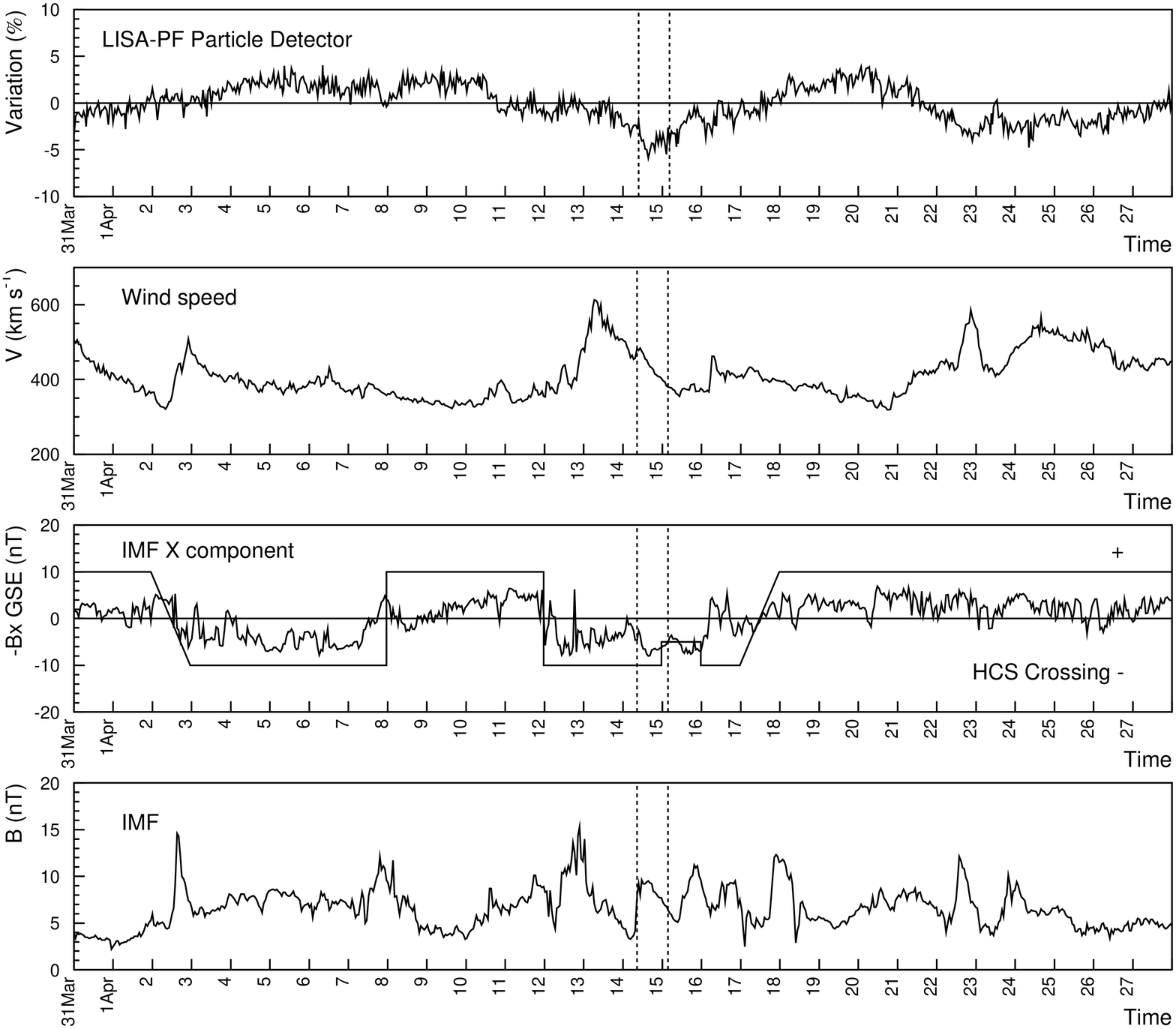}
  \caption{Same as Figure 4 for the BR 2492 (2016 March 31 - 2016 April 26).
}
  \label{figure1}
 \end{center}
\end{figure}


\section{GCR flux non-recurrent variations $<$ 2 days during LPF}

Data visual inspection of the whole LPF data set revealed the presence of several non-recurrent substructures in the GCR flux of duration shorter than two days. A dedicated analysis was carried out to investigate  the characteristics and the origin of these variations. GCR flux depressions and peaks of duration longer than 0.75 days (18 hours) with intensities $>$ 2\% were studied. GCR flux variations larger than 2\%  in intensity were considered in order to set the statistical significance of the selection criterion to  2-$\sigma$, being of 1\% the statistical uncertainty on PD hourly averaged single count data.
The LPF PD observations during each BR were compared to the IMF intensity,
solar wind plasma parameters  and  NM measurements. 
Twenty-three, non-recurrent $<$ 2-day duration  GCR flux variations were observed between 2016 February 18 and 2017 July 3. These twenty-three variations consisted of   six enhancements and seventeen depressions.  
As an example, in Figure 11 data gathered during the BR 2492 present a  small depression  on 2016 April 7-8 and two small peaks on 2016 April 15 and April 23. 
A comparison of the  LPF data with those gathered with polar NMs during the same BR 2492  in Figure 13 shows that  the small  GCR flux enhancement dated  April 15 was observed in the most of the polar NM measurements, similarly  the depression  dated 7-8 April is observed by  the Thule and McMurdo NMs. Conversely the  April 23 enhancement is not observed in polar NMs. Interplanetary plasma (solar wind bulk speed, temperature and proton density) and magnetic field parameters are studied to identify interplanetary structures associated with individual $<$2-day  GCR flux variations. 
In Table 4  CHSS, observed during subsequent Bartels rotations and originating from coronal holes, are characterized by a solar wind speed $>$ 400 km s$^{-1}$, low magnetic field and plasma densities. Corotating interaction regions (CIR) are identified as regions of compressed plasma formed between the leading edges of CHSS  at the interface that separates slow and fast stream plasma. Magnetic barriers (MBs)  indicate  those regions of high plasma magnetic field intensity observed between closely spaced CHSS. MFE stays for magnetic field enhancements  in the slow solar wind. 
The majority of small depressions in the GCR flux are  caused by  HCSC; only seldom their evolution was modulated by  CHSS and CIR.  These findings resulted to be different from  those obtained with  an analogous study carried out in \citet{armano18a} for GCR flux recurrent depressions $>2$ days indicating that, in general,  these depressions are  associated with  CIR and with the passage of CHSS.  Peaks with duration $<$ 2 days appear associated with  regions of compressed plasma  between two CHSS 
(see for instance 2016 April 23-24 in Figure 12). 
Several processes may generate these small peaks in the GCR flux. The most plausible is that the lowest energy GCRs ($\simeq$ 70 MeV) are excluded from regions of enhanced  IMF intensity between subsequent CHSS. However, a  change of the low-energy GCR spectrum slope between  flux recovery phase after a CHSS passage  and a new GCR flux decrease due to the  passage of a subsequent CHSS may also generate a peak feature in the integral flux.
An increase of the GCR flux due to the  acceleration  at the shock of  incoming CHSS does not appear plausible  on the basis of the absence of  small peak structures at the passage of  isolated CHSS (see Figure 6 in \citet{armano18a}). As a matter of fact, both models and observations indicate that the maximum energy of particles accelerated  at CIR regions is about 20 MeV \citep{mcd75,bs76,t82,desai,giacal,richar,lau15}. 

\begin{figure}
\begin{center}
  \includegraphics[width=0.7\textwidth]{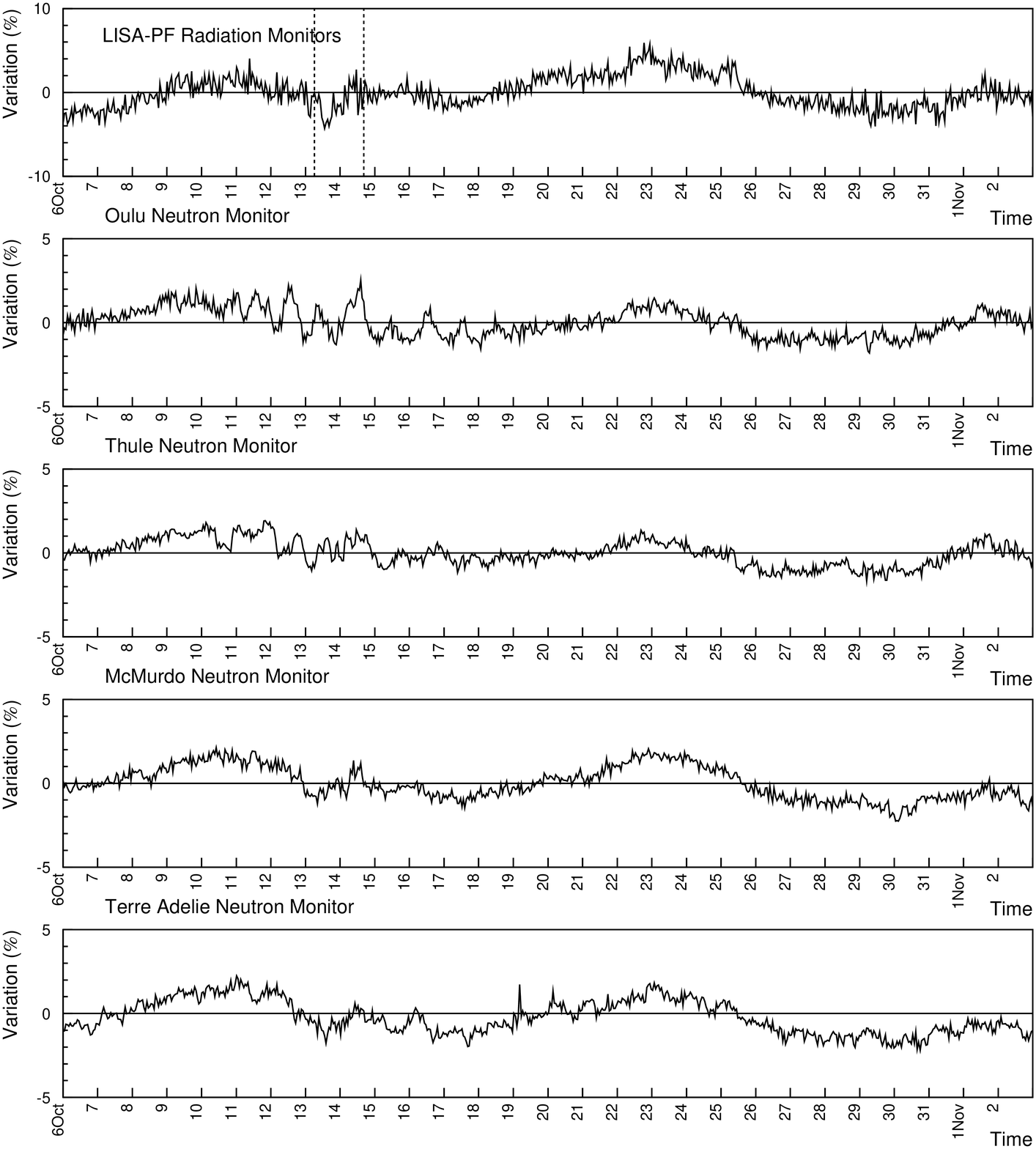}
  \caption{Comparison of LPF hourly-averaged GCR counting rate PC (top panel) with contemporaneous, analogous measurements of polar NMs during the BR 2499 (2016 October 6 - 2016 November 1). The passage of a near-Earth ICME  is indicated by vertical dashed lines (http://www.srl.caltech.edu/ACE/ASC/DATA/level3/icmetable2.htm).}
  \label{figure1}
\end{center}
 \end{figure}

\begin{figure}
  \begin{center}
  \includegraphics[width=0.7\textwidth]{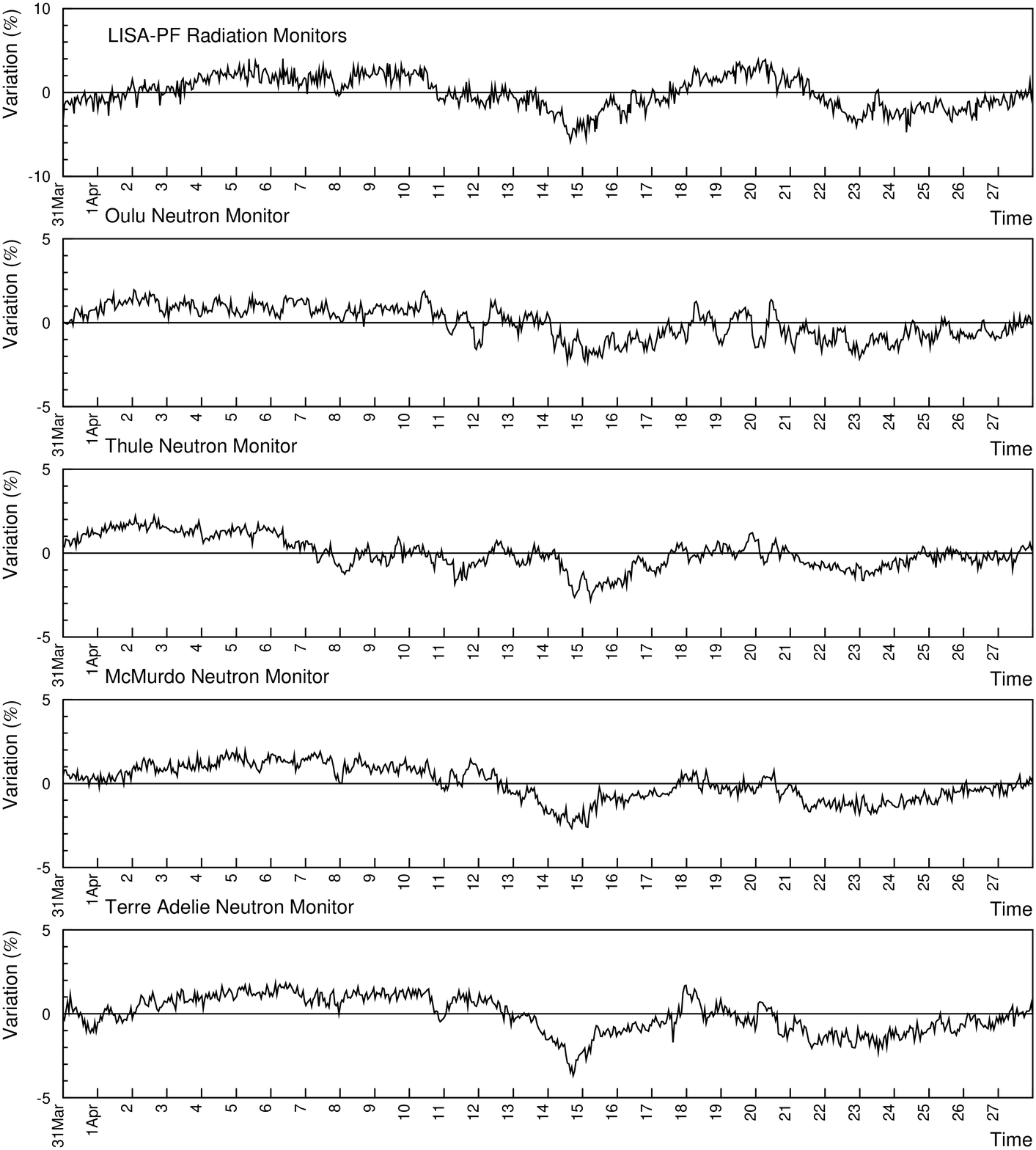}
  \caption{Same as Figure 13 for the  BR 2492 (2016 March 31 - 2016 April 26).}
  \label{figure1}
 \end{center}
\end{figure}

\startlongtable
\begin{deluxetable}{c|c|c|c|c|c}
\tablecaption{Occurrence and characteristics of the GCR flux variations $<$ 2 days observed with LPF \label{tab:table}. Interplanetary structures associated with each GCR flux  $<$ 2-day variation are indicated (CIR: corotating interaction region; CHSS: corotating high-speed solar wind streams; 
HCSC: heliospheric current sheet crossing; MFE: magnetic field enhancement in the slow solar wind; MB: low-energy cosmic rays confined in regions of high magnetic field between two subsequent CHSS). IMF, solar wind plasma data and near-earth ICME passages were gathered from the websites \url{https://cdaweb.sci.gsfc.nasa.gov/index.html}  and \url{http://www.srl.caltech.edu/ACE/ASC/DATA/level3/icmetable2.htm}. HCSC are reported in \url{http://omniweb.sci.gsfc.nasa.gov./html/polarity/polarity\_tab.html}.}
\tablehead{
  \colhead{Date} & \colhead{Onset} & \colhead{Duration} & \colhead{Dip/Peak} & \colhead{Amplitude}& Interplanetary structure\\
\colhead{} & \colhead{Time} & \colhead{Days} & & \% &
}
\startdata
2016 March 11 & 7.44 UT & 0.97 & DIP & 3.1 & CIR+HCSC\\
2016 March 16 & 1.38 UT & 0.84 & PEAK& 3.0 & MB\\
2016 March 19 & 9.21 UT & 1.16 & DIP & 2.5 & CHSS\\
2016 April 7 &  16.31 UT  & 0.89 & DIP & 3.4 &  HCSC\\
2016 April 15 &  7.44 UT  & 0.75 & PEAK & 4.4 &  MB\\
2016 April 23 &  1.13 UT & 1.16 & PEAK& 3.0 &  MB\\
2016 June 16 &  0.15 UT  & 1.74 & PEAK & 2.5 & MB\\
2016 June 21 &  23.11 UT  & 1.68 & DIP & 2.5 & HCSC\\
2016 June 23 & 2.12 UT   & 1.95 &  DIP &    2.5 &  CIR\\
2016 June 25 &  6.55 UT  & 1.79 & DIP & 2.5 &  CHSS\\
2016 June 30 & 7.19 UT   & 1.79 & DIP & 2.8 &  HCSC\\
2016 July 2 & 3.15 UT    & 2.00 & DIP & 2.8 & MFE\\
2016 July 4 & 8.57 UT    & 1.74 & DIP & 2.5 & CHSS+HCSC\\
2016 August 9 & 00.30 UT  & 0.75 & PEAK & 4.4 & MB\\
2016 August 16 & 6.06 UT & 1.79 & DIP & 2.5 &  HCSC\\
2016 September 15 & 22.43 UT & 0.84 & DIP& 3.1 &  MFE\\
2016 October 11 &  13.25 UT   & 0.95 & DIP & 2.5 &  CHSS\\
2016 October 23 & 21.09 UT    & 0.95 & DIP & 3.8 &  HCSC+CHSS\\
2016 November 10 &  17.29 UT  & 1.16 & DIP & 3.8 & CIR+HCSC\\
2017 January 14 &  12.52 UT   & 0.75 & DIP &  3.1& HCSC\\
2017 March 22 & 22.21 UT   & 1.05 & PEAK & 2.1 &   MB\\
2017 May 5&  12.17 UT & 1.46 & DIP & 2.3 & MFE\\
2017 May 29 & 21.07  UT  & 1.53 & DIP & 2.4 &  CIR\\
\enddata
\end{deluxetable}

\section{Conclusions}
A PD aboard LPF allowed for the measurement of the integral flux variation of GCR protons and helium nuclei above 70 MeV n$^{-1}$.
The energy-dependence of FDs measured aboard LPF was compared to that of other space experiments and  NMs which allow for a direct measurements of the 
integral flux variation of GCRs above effective energies $>$  10 GeV. A  parameterization  of  pre-decrease energy spectra  
 and energy spectra measured during the 
main phase of  FDs is found to apply to different intensity  events.

FDs observed in L1 
are not correlated with geomagnetic storm occurrence  unless 
the southward component (B$_z$) of the interplanetary magnetic field  presents values $<$  -20 nT.
Finally,  hourly averaged GCR flux variations measured with LPF allowed for the observations of non-recurrent features in the GCR integral flux variations $>$ 0.75 days and $<$ 2 days with  intensities $>$ 2\%. These short-term depressions and peaks in the data trend appear correlated in the majority of cases with HCSC and plasma compression regions between subsequent CHSS, respectively.       




\acknowledgments
This work is dedicated to the loving memory of Prof. Karel Kudela, a marvelous colleague and a special friend that left us too soon on January 20 2019.

The authors are grateful to the anonymous referee for his/her precious comments and suggestions that allowed for a major improvement of the manuscript.

Solar modulation parameter data were gathered from 
 \url{http://cosmicrays.oulu.fi/phi/Phi\_mon.txt}. Data from the ACE experiment were obtained from the NASA-CDAWeb website. We acknowledge the NMDB database (www.nmdb.eu) funded under the European Union's FP7 programme (contract no. 213007), and the PIs of individual NM stations for providing data.  HCS crossing was taken from \url{http://omniweb.sci.gsfc.nasa.gov./html/polarity/polarity\_tab.html}.

This work has been made possible by the LISA Pathfinder mission, which is part of the space-science program of the European Space Agency. The French contribution has been supported by the CNES (Accord Specific de projet CNES 1316634/CNRS 103747), the CNRS, the Observatoire de Paris, and the University Paris-Diderot. E. P. and H. I. also acknowledge the financial support of the UnivEarthS Labex program at Sorbonne Paris Cit\'e (ANR-10-LABX-0023 and ANR-11-IDEX-0005-02). The Albert-Einstein-Institut acknowledges the support of the German Space Agency, DLR. The work is supported by the Federal Ministry for Economic Affairs and Energy based on a resolution of the German Bundestag (FKZ 50OQ0501 and FKZ 50OQ1601). The Italian contribution has been supported by Agenzia Spaziale Italiana and Instituto Nazionale di Fisica Nucleare. The Spanish contribution has been supported by Contracts No. AYA2010-15709 (MICINN), No. ESP2013-47637-P, and No. ESP2015-67234-P (MINECO). M. N. acknowledges support from Fundacion General CSIC (Programa`ComFuturo). F.R. acknowledges an FPI contract (MINECO). The Swiss contribution acknowledges the support of the Swiss Space Office (SSO) via the PRODEX Program of ESA. L. F. acknowledges the support of the Swiss National Science Foundation. The United Kingdom groups acknowledge support from the United Kingdom Space Agency (UKSA), the University of Glasgow, the University of Birmingham, Imperial College, and the Scottish Universities Physics Alliance (SUPA). J. I. T. and J. S. acknowledge the support of the U.S. National Aeronautics and Space Administration (NASA). N. Korsakova acknowledges the support of the Newton International Fellowship from the Royal Society. KK was formerly supported by the project CRREAT (reg. CZ.02.1.01/0.0/0.0/15003/0000481)
call number 02 15 003 of the Operational Programme Research, Development and
Education.

\end{document}